\documentclass[10pt,preprint]{aastex}
\usepackage{rotating}
\usepackage{array}

\begin{document}

\title{Infrared Dark Clouds in the Small Magellanic Cloud?}

\author{Min-Young Lee\altaffilmark{1}, 
Sne\v{z}ana Stanimirovi\'c\altaffilmark{1}, 
J\"urgen Ott\altaffilmark{2},  
Jacco Th. van Loon\altaffilmark{3}, 
Alberto D. Bolatto\altaffilmark{4}, \\
Paul A. Jones\altaffilmark{5,6}, 
Maria R. Cunningham\altaffilmark{5}, Kathryn E. Devine\altaffilmark{1}, 
and Joana M. Oliveira\altaffilmark{4}}

\altaffiltext{1}{Department of Astronomy, University of Wisconsin-Madison, 
475 North Charter Street, Madison, WI 53706; lee@astro.wisc.edu, 
sstanimi@astro.wisc.edu, and devine@astro.wisc.edu} 
\altaffiltext{2}{National Radio Astronomy Observatory, 520 Edgemont Road, 
Charlotteville, 
VA 22903; jott@nrao.edu}
\altaffiltext{3}{Astrophysics Group, Lennard Jones Laboratories, 
Keele University, 
Staffordshire, ST5 5BG, UK; jacco@astro.keele.ac.uk, and joana@astro.keele.ac.uk}
\altaffiltext{4}{Department of Astronomy and Laboratory for Millimeter-wave 
Astronomy, University of Maryland, College Park, MD 20742; bolatto@astro.umd.edu}
\altaffiltext{5}{School of Physics, University of New South Wales, Sydney, 
UNSW 2052, Australia; pjones@phys.unsw.edu.au, and 
maria.cunningham@unsw.edu.au}
\altaffiltext{6}{Departamento de Astronomia, Universidad de Chile, Casilla 36-D, 
Santiago, Chile}
 
\begin{abstract}
We have applied the unsharp-masking technique to the 24 $\mu$m image of 
the Small Magellanic Cloud (SMC), obtained with the \textit{Spitzer Space Telescope}, 
to search for high-extinction regions.
This technique has been used to locate very dense and cold 
interstellar clouds in the Galaxy, particularly infrared dark clouds (IRDCs).
Fifty five candidate regions of high-extinction, namely high-contrast regions (HCRs),  
have been identified from the generated decremental contrast image of the SMC. 
Most HCRs are located in the southern bar region and 
mainly distributed in the outskirts of CO clouds, but most likely contain a 
significant amount of H$_{2}$.
HCRs have a peak-contrast at 24 $\mu$m of $2 - 2.5$ \% and a size of $8 - 14$ pc. 
This corresponds to the size of typical and large Galactic IRDCs,
but Galactic IRDCs are $2 - 3$ times darker at 24 $\mu$m than our HCRs.
To constrain the physical properties of the HCRs, 
we have performed NH$_{3}$, N$_{2}$H$^{+}$, 
HNC, HCO$^{+}$, and HCN observations toward one of the HCRs, 
HCR LIRS36--EAST, using the Australia Telescope Compact Array 
and the Mopra single-dish radio telescope. 
We did not detect any molecular line emission, however, our upper limits to 
the column densities of molecular species suggest that HCRs are most likely 
moderately dense with $n \sim 10^{3}~\rm cm^{-3}$. 
This volume density is in agreement with predictions for the cool atomic 
phase in low metallicity environments. 
We suggest that HCRs may be tracing clouds at the 
transition from atomic to molecule-dominated medium,
and could be a powerful way to study early stages of gas condensation 
in low metallicity galaxies.
Alternatively, if made up of dense molecular clumps $<0.5$ pc in size, HCRs could be 
counterparts of Galactic IRDCs, and/or regions with highly unusual abundance
of very small dust grains.
\end{abstract}

\section{Introduction}

The discovery of infrared dark clouds (IRDCs) 
through the \textit{Infrared Space Observatory (ISO)} 
and the \textit{Midcourse Space Experiment (MSX)} surveys of the Galactic plane 
(P\'erault et al. 1996; Carey et al. 1998; Egan et al. 1998)
opened a new window for studies of the dense and cold interstellar medium (ISM).
IRDCs, a new class of interstellar clouds, are seen as dark extinction features 
against the bright Galactic background at mid-infrared (mid-IR) wavelengths. 
In-depth studies with images obtained by the \textit{Infrared Astronomical 
Satellite (IRAS)} and the \textit{MSX} showed that IRDCs are opaque
at wavelengths from 7 to 100 $\mu$m (Egan et al. 1998). 
Millimeter and submillimeter observations confirmed that 
IRDCs are dense ($n > 10^{5}~\rm cm^{-3}$) and cold ($T_{gas} < 20$ K),
and have high H$_{2}$ column densities ($10^{23-25}~\rm cm^{-2}$) 
(Carey et al. 1998, 2000; Rathborne et al. 2005). 

A new approach for studying the dense and cold ISM, based on 
the unsharp-masking technique, may be particularly useful for 
low metallicity environments where, due to a low shielding ability, 
molecules can survive only in dense cores  (Elmegreen 1989)
and thus it is extremely difficult to find them with blind searches. 
High-sensitivity and high-resolution IR images, 
obtained with the \textit{Spitzer Space Telescope} 
(\textit{Spitzer} throughout this paper; Werner et al. 2004), are 
suitable for searching for dark clouds 
at mid-IR wavelengths to locate potential dense molecular clouds.
Motivated by the recent \textit{Spitzer} observations of the Small Magellanic Cloud (SMC) 
(Bolatto et al. 2007), we undertook the first search for IR dark clouds in the SMC.

One of the closest neighbors of the Galaxy, 
the SMC is a low mass ($M_{dyn} \sim 2.4\times 10^{9}~\rm M_{\odot}$; 
Stanimirovi\'c et al. 2004), gas-rich (Staveley-Smith et al. 1998), 
dwarf irregular galaxy at a distance of approximately 60 kpc\footnote{
Throughout this paper we assume a distance to the SMC of 60 kpc.} 
from Earth (Westerlund 1991). 
Like its companion galaxy, the Large Magellanic Cloud (LMC), 
the SMC has a number of active star-forming regions.  
Most of the star formation activity in the SMC is concentrated 
in the feature known as the bar. 
The SMC bar contains most of the dense gas as well. 
Rubio et al. (1991) pointed out the difference in the star-formation activity
along the bar, with the south-west (SW) region being more active and containing 
a large number of young objects which correlate well with CO clouds, while
the north-east (NE) region containing only smaller CO clouds with almost no association with
young stellar objects. 
The second prominent large-scale feature in the SMC, the wing, is 
located to the south-east (SE) of the SMC and 
contains young stellar components (e.g. N83/N84 region) 
which imply recent star formation
(Westerlund \& Glaspey 1971; Kunkel 1980).  

The interstellar environment in the SMC is very different 
from that in the Galaxy. 
The SMC has a much lower heavy element abundance 
($Z \sim 0.2~\rm Z_{\odot}$;  Dufour 1975) and an interstellar UV radiation field (ISRF)
$4-10$ times higher than that in the Solar neighborhood 
(Lequeux 1979; Vangioni-Flam et al. 1980; Azzopardi et al. 1988). 
Hence, the SMC provides a unique laboratory to study how 
different galactic interstellar environments affect properties and evolution of the ISM.
In addition, the SMC is the closest local example
of a relatively primitive ISM and provides a stepping 
stone in our understanding of the ISM in high-redshift galaxies.

While many authors have investigated properties of molecular gas in the SMC
based on CO observations, only a few measurements of molecular 
species other than CO exist (Chin et al. 1998; Heikkil\"a et al. 1999).
Rubio et al. (1993) and Lequeux et al. (1994) analyzed CO data obtained 
by the Swedish-ESO Submillimeter Telescope (SEST) and found that 
CO(1--0) emission is deficient in the SMC by a factor of $10 - 20$ relative to the Galaxy. 
Their results also showed that CO clouds in the SMC are clumpy 
with a surface filling factor lower than for Galactic clouds. 
In addition, several studies suggested that the CO emission 
is found only in dense cores, and that molecular hydrogen 
(H$_{2}$) may have a more extended distribution.
Mizuno et al. (2001) compared CO luminosities for clouds
observed by both NANTEN ($\rm angular~resolution \sim 2.6'$) 
and SEST ($\rm angular~resolution \sim 45''$) 
and concluded that the CO emission is highly clumped, with
no diffuse CO emission in the inter-cloud medium. 
Rubio et al. (2000) performed observations of 
the $v$=(1--0)S(1) line of H$_{2}$ toward N66 and 
found that CO molecules only exist in regions with 
$A_{V}$ larger than 1 mag, compared to 0.1 mag or less for the 
the $v$=(1--0)S(1) line of H$_{2}$. 
Using the \textit{Spitzer} 24, 70, and 160 $\mu$m images, 
Leroy et al. (2007) constructed the H$_{2}$ surface density map
under the assumption that all excess of far-IR emission, relative to
the observed H \textsc{i} emission (corrected for self-absorption), 
comes from the H$_{2}$ component.
They estimated $M_{\rm H_{2}}=3.2 \times 10^7~M_{\odot}$ and 
suggested that H$_{2}$ is on average $1 - 3$ times more extended than CO
\footnote{It is important to point out that the excess far-IR emission
could result from an increased far-IR emissivity, 
possibly due to growth of icy mantles on the surface of dust grains, and/or
spatial variations of the dust-to-gas ratio.}.

Several studies have used IR emission to explore properties of dust in the SMC. 
Sauvage et al. (1990) studied the effects of stellar age and metallicity 
on the IR emission in the Magellanic Clouds and found that 
there is a decrease of the 12 $\mu$m flux as a function of metallicity, 
which indicates a relative under-abundance of the polycyclic aromatic hydrocarbons 
(PAHs) compared to the large grain population. 
Similarly, Stanimirovi\'c et al. (2000) suggested that PAH molecules and 
very small grains (VSGs) contribute little to the dust emission at wavelengths 
longer than 45 $\mu$m, based on \textit{IRAS} images of the SMC.
On the other hand, the recent LMC study with the Surveying the Agents of a 
Galaxy's Evolution (SAGE) data suggested that the 70 $\mu$m excess observed 
in the spectral energy distributions (SEDs) of both the LMC and SMC could be due to the 
abundant large VSGs produced through erosion of larger grains in the diffuse medium 
(Bernard et al. 2008). 

The main goal of this study is to search for high-extinction regions at mid-IR wavelengths 
in the SMC by applying the unsharp-masking technique.
By obtaining further molecular line observations, we aim at constraining
physical properties of these regions.
The \textit{Spitzer} 24 $\mu$m image used for this study was obtained 
as part of the $\rm S^{3}MC$ survey (Bolatto et al. 2007) and
has an excellent spatial resolution (1.7 pc) high enough to disentangle individual 
molecular clouds.

This paper is organized as follows. 
In Section 2 we use the \textit{Spitzer} 24 $\mu$m image of the SMC to generate
the decremental contrast image and search for high-extinction regions 
(high-contrast regions; HCRs throughout this paper). 
In Section 3 overall properties of identified HCRs are investigated. 
In Section 4 we examine characteristics 
of several selected HCRs using the available multi-wavelength data. 
In Section 5 we present our observations of $\rm NH_{3}$, 
$\rm N_{2}H^{+}$, HNC, $\rm HCO^{+}$, and HCN 
toward one HCR (named as HCR LIRS36--east)
and derive upper limits to the column densities of observed molecular species.   
Constraints on the physical properties and potential origins of HCRs are discussed in Section 6. 
We summarize our main findings in Section 7. 

\section{Search for High-Extinction Regions in the SMC: 
The 24 $\mu$m Contrast Image}

In the Galaxy, high-extinction regions (like IRDCs) are identified as silhouettes 
against the diffuse Galactic background at mid-IR wavelengths that easily stand out
when an unsharp-masking technique is applied (e.g. Simon et al. 2006a).
Assuming that similar reasoning will apply in the SMC, 
we have searched for fractional decrements in the \textit {Spitzer} 24 $\mu$m image.
We followed the method of Simon et al. (2006a) who identified and cataloged 
10,931 Galactic IRDC candidates by searching for contiguous regions of 
high-contrast with respect to the background of the \textit{MSX} 8.3 $\mu$m image. 
Since the small dust grains, especially PAHs are less abundant in the SMC than in 
the Galaxy (Sauvage et al. 1990; Stanimirovi\'c et al. 2000), 
the 8 $\mu$m image of the SMC obtained with the \textit{Spitzer} (Bolatto et al. 2007) 
does not show significant diffuse emission.
In addition, the diffuse 8 $\mu$m emission is mainly concentrated around H \textsc{ii}
regions, providing very little diffuse IR background for unsharp-masking. 
The diffuse IR emission at 24 $\mu$m is more pronounced, 
providing the opportunity to look for local spatial variations.
In addition, at least one 
sharp silhouette is 
noticeable by eye in the southern bar region, 
giving us confidence that this method may reveal some interesting structures.

The \textit{Spitzer} 24 $\mu$m image has an angular resolution of 6$''$
and this corresponds to a linear size of 1.7 pc at a distance of 60 kpc. 
The 24 $\mu$m wave band is sensitive to continuum emission from PAHs and VSGs 
($a \lesssim 0.005$ $\mu$m) but much of the 24 $\mu$m emission in the SMC
is expected to come from stochastically heated ($\sim 150$ K) VSGs. 
For example, Madden et al. (2006) showed that the mid-IR emission from N66 
is dominated by VSGs. 
These authors used one of ISOCAM bands, covering the wavelength range 
from 12 to 18 $\mu$m, and estimated that $85 - 95$ \% of the IR flux 
comes from VSGs. 
To search for high-extinction regions,
we first determined the diffuse IR background at 24 $\mu$m and then derived the
decremental contrasts against this background.

\subsection{Determining the Diffuse 24 $\mu$m Background}

We assumed that the \textit{Spitzer} 24 $\mu$m image,
apart from bright emission from star-forming regions and point sources, 
consists of the diffuse IR background against which high-extinction clouds are superposed. 
To determine the background, we smoothed the \textit{Spitzer} 24 $\mu$m image 
using the technique of spatial median filtering. 
The median filtering size must be larger than the typical size of high-extinction regions,
but small enough to follow the spatial distribution of the background. 
Typical IRDCs in the Galaxy are smaller than 10 pc 
(or 0.6$'$, which is $\sim 10$ pixels in the 24 $\mu$m image;
one pixel corresponds to $1 \times 10^{-3}$ deg), 
and we used this as the absolute minimal size for our smoothing kernel.
On the other hand, as the spatial variation of the background at 24 $\mu$m is 
typically larger than 100 pc (or 6$'$, which is $\sim 95$ pixels in the 24 $\mu$m image), 
the median filtering size must be smaller than 95 pixels. 
We examined the spatial variation of the background at 24 $\mu$m 
by obtaining profiles across the \textit{Spitzer} image at various positions. 
We varied the median filtering size from about 33 to 95 pixels and 
selected the kernel with 63 pixels (or 66 pc) as providing the best estimate of
the IR background. 

Figure 1 shows the estimated 24 $\mu$m background image (top) and a profile 
along the $x$-axis of the background image superposed on the original 
24 $\mu$m data (bottom).
For the median-kernel with 63 pixels, the diffuse background emission contributes only 
27 \% of the total 24 $\mu$m emission. 
Most of the 24 $\mu$m emission comes from H \textsc{ii} regions, in particular, 
those in the southern bar region. 
The ratio of diffuse to total 24 $\mu$m emission varies from 23 \% in the NE portion 
of the bar, to 31 \% in the SW portion of the bar. 

\subsection{The Contrast Image}

After determining the background, we generated the contrast image by subtracting 
the \textit{Spitzer} 24 $\mu$m image from the estimated background and 
then dividing the residual image by the background image, 
$\rm contrast=(background-original~image)/background$. 
By this definition, regions having lower surface brightness against the background 
will have positive contrasts. 
The H \textsc{ii} regions have negative contrasts as
their IR emission is higher than the IR background and we exclude 
them from further analysis.
Our detection method is therefore limited to positive contrasts and 
we are essentially blind to high-extinction regions that may be present within 
the boundaries of H \textsc{ii} regions. 

The (positive) contrasts are typically very small and range from 0 to 4 \%.
To estimate the significance of contrasts, we derived the
signal-to-noise ratio image. The signal-to-noise ratio here is defined by the ratio 
of the contrast to its error ($c/\Delta c$). 
The contrast error ($\Delta c$) for a particular pixel was calculated by dividing the average 
$1\sigma$ noise level in the 24 $\mu$m image 
by the pixel value in the background image.
We have assumed a constant value for the 
$1\sigma$ noise level of 0.06 $\rm MJy~sr^{-1}$ by considering 
several emission-free regions in the image. 
The derived signal-to-noise ratio image is shown in Figure 2.

Figure 3 shows a histogram of pixel values 
from the signal-to-noise ratio image for the SMC SW bar.
For small values ($<3\sigma$), this distribution can be well fit with a Gaussian
function, suggesting that a majority of pixels have low significance and are
purely due to Gaussian noise.
However, the distribution has a significant tail, peaking around 4$\sigma$, and suggesting
that a certain fraction of pixels with high-significance deviate from the Gaussian 
statistics and could represent real features.
 
\subsection{Selection of High-Contrast Regions (HCRs)}

Figure 2 shows many small and large features with significant
fractional decrements of the 24 $\mu$m intensity relative to the IR background.
Figure 3 shows a significant fraction of pixels with $c>3\sigma$.
To select reliable HCRs, we take a conservative approach and search for contiguous regions with
contrasts higher than 5$\sigma$.
This is mainly motivated by a high likelihood of spurious detections at low contrasts  
(discussed below). 
In addition, we require HCRs to be extended and contain at least $9 - 10$ pixels.  
This corresponds to twice the full width half maximum (FWHM) beam solid angle
of the \textit{Spitzer} at 24 $\mu$m ($2~\Omega = 2.26\times(6'')^{2} = 81~\rm
arcsec^{2} \sim 3\times 3~pixels$). 
Therefore, the smallest clouds we are sensitive to have a size of $\sim3$ pc.
This is similar to the typical size of Galactic IRDCs, $\sim 4$ pc (Simon et al. 2006b), 
although Galactic IRDCs can range in size from 0.4 to 10 pc 
(Pillai et al. 2006; Frieswijk et al. 2007).
We are therefore sensitive to the counterparts of typical and large Galactic IRDCs.
By these criteria, 55 candidate HCRs (39 from the southern bar, 
6 from the eastern wing, and 10 from the northern bar) were identified. 

However, it is possible that some HCRs are spurious and do not
trace regions of high-extinction due to the following 
limitations of our method.

(1) The HCRs have peak-contrasts up to 4 \%, which is lower
than what is found for Galactic IR images (see Section 3 for details). 
For the Galactic HCRs, Simon et al. (2006a) and Jackson et al. (2008) showed that 
higher contrast candidates have high CS(2--1) line 
($n_{crit}=7\times10^{5}~\rm cm^{-3}$) detection rates 
and are therefore prime regions for being real IRDCs. 
For contrasts of $c_8=20-40$ \% at 8 $\mu$m, which corresponds to
$c_{24}=4-10$ \% at 24 $\mu$m (assuming the relation $A_{\lambda} \propto \lambda^{-1.7}$ 
in the near-IR regime; Mathis 1990), their detection rate of CS(2--1) emission was 50 \%, 
while for $c_8> 60$ \%, which corresponds to $c_{24}>20$ \%,  
the detection rate increased to nearly 100 \%. 
Based on these studies, HCRs with $c_{24}<4$ \% are
expected to have high likelihood of being spurious detections in terms of being tracers
of regions with the highest extinction.

(2) To be detected, HCRs need to lie in front of the bulk of diffuse IR emission. 
The SMC has been postulated to have a significant line-of-sight depth 
(e.g. Mathewson et al. 1986; Welch et al. 1987), 
which complicates the detection of HCRs. 
HCRs in the parts of the SMC with lower line-of-sight depth (NW)
and at a near distance will be easier to find, 
while HCRs in regions with a larger depth (SW) will tend to be filled by the 
foreground emission and hard to detect (see Section 3 for details). 

(3) The uncertainty and fluctuations in the IR background, 
especially close to complex H \textsc{ii} regions, affect the reliability of HCRs. 
It has been pointed out by several studies that the unsharp-masking technique 
cannot distinguish voids in regions of complex emission from actual extinction features. 
This results in false detections and the exact reliability of the method is best 
constrained through observations of molecular lines (Jackson et al. 2008).

After taking into consideration all selection effects, 
we stress that our detection method is blind to high-extinction regions 
within the boundaries of H \textsc{ii} regions and is biased towards the outskirts of 
H \textsc{ii} regions where the diffuse IR emission is stronger than in outer parts of the SMC. 
Considering that HCRs with very high-contrasts in the Galaxy show 
a high detection rate in molecular line observations,
we focus in this paper primarily on several selected regions
with the highest contrasts, and therefore the highest likelihood of being real.
The best approach for checking the reliability of HCRs is through a detection
and morphological matching of molecular line emission/continuum data which we attempt
in Section 4 and 5.

\section{Overall Properties of HCRs}

Figure 2 shows the spatial distribution of HCRs, 
with most of them (39 out of 55) being in the SW portion of the SMC bar.
Other SMC regions contain only a few HCRs. 
While the exact line-of-sight depth of the SMC is uncertain, 
most studies agree that the depth is higher in the SW portion of the bar 
than in the NE region (Mathewson et al. 1986; Groenewegen 2000;
Subramanian \& Subramaniam 2009).
Based on our selection bias discussed in Section 2.3, 
this effect would result in a smaller number of HCRs in the SW
region, yet we see an opposite trend. 
The distribution of H \textsc{ii} regions seen in Figure 2 (as negative, white pixels)
is relatively uniform along the bar; no
strong north-south asymmetry is observed.
In addition, the fraction of IR background emission relative to the total emission 
is not very different between the NE and the SW,
23 \% versus 31 \% respectively.
This all suggests that the highly inhomogeneous spatial distribution of HCRs,
with a large abundance of HCRs in the SW, is not a selection bias 
but is an intrinsic property.

It is interesting to note that the large disproportion in the 
spatial distribution of HCRs is qualitatively
similar to the spatial distribution of CO clouds by Mizuno et al. (2001).
Most of the CO clouds (ranging in size from 50 to 160 pc) are located in 
the SW bar, with only a few sporadic clouds being found elsewhere.
The total area of CO clouds corresponds to $\sim5$ \% for the NE 
and $\sim11$ \% for the SW (Mizuno et al. 2001). 
Young emission objects (e.g. H \textsc{ii} regions with embedded 
OB associations and star clusters) tracing star formation in the range 
$<5-6$ Myrs were found to be well correlated
with CO clouds, while emissionless objects 
(e.g. emissionless OB associations and star clusters of ages $\sim 6-100$ Myrs) 
do not show any particular association with CO clouds.
This suggests that HCRs are intimately connected with CO clouds and
active (more recent) sites of star formation.

In Figure 4 we present two histograms showing the peak-contrast and
size distribution of selected 55 HCRs. 
The peak-contrast of HCRs ranges from 1.7 to 3.5 \% and 
peaks at $2 - 2.5$ \%. 
The distribution of peak-contrasts of Galactic IRDCs at 8 $\mu$m is approximately 
a power-law, with a peak around $15 - 20$ \% (full range $10 - 60$ \%, 
Simon et al. 2006a).
After converting this to the expected extinction at 24 $\mu$m 
(using the relation $A_{\lambda} \propto \lambda^{-1.7}$ in the near-IR regime; Mathis 1990), 
HCRs in the SMC have a peak-contrast a factor of $2 - 3$ lower than Galactic IRDCs. 

The size distribution in Figure 4 (bottom) shows that 80 \% of the  
selected HCRs have a size in the range from 3 to 18 pc 
(to estimate cloud size we counted the number of pixels within the HCRs and 
then assumed simply spherically symmetric clouds). 
The lower limit of this size distribution is set by our selection criteria and 
resolution. There is a strong peak at $8 - 14$ pc and a tail going 
all the way to $\sim50$ pc. 
Galactic IRDCs have diameters of $0.4 - 10$ pc 
(Pillai et al. 2006; Frieswijk et al. 2007), with a typical size of 
$\sim4$ pc (Simon et al. 2006b). This is much smaller than the
average diameter of giant molecular clouds (GMCs) in the Galaxy, which is about 50 pc.
HCRs in the SMC therefore, size wise, 
are significantly smaller than the GMCs in either the Galaxy or the SMC
and similar to counterparts of typical and large Galactic IRDCs. 

Under the assumption that HCRs are entirely due to the extinction of 
IR background emission at 24 $\mu$m, 
we can use the following simple radiative transfer equation to estimate their optical depth
\begin{equation}
c = \beta (1-e^{-\tau}),
\end{equation}
\noindent 
where $c$ is the peak-contrast, $\beta=I_{bh}/I_{bg}$ is the fraction
of IR emission behind the HCR relative to the total IR background emission 
along the line-of-sight, 
and $\tau$ is the optical depth at 24 $\mu$m. 
By definition $\beta \leq1$, $\beta=1$ if HCRs are at the front of the SMC 
and $\beta=0.5$ if HCRs are in the middle of the SMC. 
By assuming $\beta = 0.5$ and using the relation $A_{V} = A_{24}/0.045$ 
(based on Figure 1 and 4 of Draine et al. 2003), 
we obtained 
$A_{V} = 1 - 1.2$ mag for HCRs with $c_{24} = 0.02 - 0.025$. 
This is moderately high for the SMC, considering that molecular peaks
in the SMC have $A_{V} = 1 - 2$ mag (Leroy et al. 2007).

We have considered possible variations in $A_{V}$ when $\beta$ varies from 0.1 to 1.
For $\beta = 0.1$, the estimated $A_{V}$ is significantly higher, 
$5 - 7$ mag.
However, it would be very hard to detect HCRs from the far side of the SMC as
they would not have much IR background emission to absorb. In addition, even if
a HCR at a far distance has a suitable IR background
against which to absorb, the dark extinction feature will tend to be filled 
by the foreground IR emission along the line-of-sight.
It is therefore highly unlikely to observe HCRs with such low $\beta$ values.
On the other hand, if HCRs are at the front of the SMC with $\beta = 1$,
the estimated $A_{V}$ would be slightly lower, $0.5 - 0.6$ mag.
However, using the H \textsc{i} surface density map (Stanimirovi\'c et al. 1999), 
which has a linear resolution of 30 pc,  
and applying the relation $N$(H \textsc{i}$)/A_{V} = (13.18 \pm 1.02) \times 10^{21}$ 
established for the SMC bar region (Gordon et al. 2003), we
estimate the expected extinction of $A_{V} = 0.8 - 1$ mag in the direction of HCRs.  
This is a lower limit as denser regions are most likely smaller than
30 pc.  This suggests that $\beta = 1$ assumption most likely underestimates
the amount of extinction, and therefore we assume $\beta = 0.5$ and 
$A_{V} \sim 1$ mag throughout the paper.

Using the H$_{2}$ surface density map of Leroy et al. (2007),
we estimated H$_{2}$ surface densities of the HCRs, 
$\rm \Sigma_{H_{2}} = 60 - 180~M_{\odot}~pc^{-2}$. 
Molecular peaks in the SMC are typically found to have 
only slightly higher surface densities, 
$\rm \Sigma_{H_{2}} = 100 - 200~M_{\odot}~pc^{-2}$ (Leroy et al. 2007),
suggesting that HCRs contain a significant amount of H$_{2}$.
Interestingly, contributions from H \textsc{i} and $\rm H_{2}$ to 
the total hydrogen column density are roughly similar, 
which was also noted for the SMC as a whole by Oliveira (2009).

 

We conclude that the global properties of the HCRs suggest they are related to recent sites
of star formation, and have sizes similar or larger than those of Galactic IRDCs.
The peak-contrasts of HCRs, $A_V\sim1$ mag, and the estimated H$_2$ 
surface density suggest regions of moderately high-density.

\section{Selected HCRs}

In this section we focus on several of the most prominent HCRs and examine
their properties based on the available multi-wavelength observations.
 
\subsection{HCRs in the N83/N84 Region}

N83 and N84 are isolated, relatively active star-forming regions in the SMC wing. 
The top panel of Figure 5 shows the CO(2--1) integrated intensity map 
from Bolatto et al. (2003) overlaid on the Digitized Sky Survey (DSS) R band image 
of the N83/N84 region. 
In the bottom panel of Figure 5 selected HCRs are shown 
as black contours on the background of CO(2--1) intensity map. 
Four large features with the highest contrasts are labeled.  
The low abundance of stars between N83 and N84 is clearly seen 
in the DSS R band image, which suggests the existence of high-extinction gas. 
Interestingly, 4 HCRs (size range $6 - 21$ pc), weakly traced by CO(2--1) emission, 
are identified exactly in this region.  

Bolatto et al. (2003) found two distinct regions with anomalously 
high ratios of CO(2--1)/CO(1--0) (N83 and N84D in Figure 1 of Bolatto et al. 2003). 
They suggested that this CO emission arises 
from an ensemble of small ($R \sim 0.1$ pc), 
moderately dense ($n \sim 10^{4}~\rm cm^{-3}$), 
and fairly warm ($T \sim 40$ K) clumps. 
The 4 HCRs are located between the two regions with high ratios of CO(2--1)/(1--0).
This suggests that HCRs may trace gas compressed between N83 and N84. 
As N83 is an expanding molecular shell, this would give a mechanism for warm CO gas 
to be swept up and compressed. 




Leroy et al. (2009) investigates the relationship between 
the integrated CO(1--0) and CO(2--1) emission
($I_{\rm CO}$) and $\Sigma_{\rm H_{2}}$ in the N83/N84 region at 10 pc resolution. 
They find that $I_{\rm CO(1-0)} = 0~\rm K~km~s^{-1}$ 
corresponds to roughly $\Sigma_{\rm H_{2}} = 50 - 150~\rm M_{\odot}~pc^{-2}$ and 
interpret this as the presence of an envelope of $\rm H_{2}$ with little or no associated CO.
Regions with enough high dust shielding ($A_{V} > 2$ mag; 
$A_{V}$ is estimated from the \textit{Spitzer} 160 $\mu$m image 
assuming optically thin dust grains)
are found to have $I_{\rm CO(1-0)} >0~\rm K~km~s^{-1}$. The
threshold of $A_{V} > 2$ mag required for CO survival in the presence of an 
intense radiation field is in agreement with Lequeux et al. (1994)'s modeling.
The 4 HCRs in this region have $\Sigma_{\rm H_{2}} = 100 - 150~\rm M_{\odot}~pc^{-2}$, 
very little CO, and 
$A_{V} = 0.6 - 1.5$ mag.
This is in agreement with properties
of CO-weak, H$_{2}$-dominated envelopes.

In summary, from the studies of Bolatto et al. (2003) and Leroy et al. (2009), 
we suggest that HCRs in the N83/N84 region most likely trace 
moderately dense gas ($n\sim10^{4}$ cm$^{-3}$ as the critical density for
CO(2--1) is $\sim10^4$ cm$^{-3}$) in the extended H$_{2}$-dominated 
molecular cloud envelope, on spatial scales of $6 - 20$ pc. 



\subsection{HCRs in the SMC--B2 Region}

The SMC--B2 region is located in the southern bar at 
($\alpha$, $\delta$)$_{\rm J2000}=
($$\rm 00^{h}$:$\rm 48^{ m}$:$\rm 01.96^{s}$, 
$\rm -73^{\circ}$:$\rm 15'$:$\rm 38.5''$).
This region contains many HCRs, whose sizes range from 4 to 48 pc. 
The top panel of Figure 6 shows the CO(1--0) integrated intensity map 
overlaid on the DSS R band image. 
In the bottom panel of Figure 6 selected HCRs are shown in black contours
with the background of CO(1--0) intensity map. 
Two large features with the highest contrasts are labeled.
HCR LIRS36--east, which will be discussed in the next subsection, is also shown in Figure 6. 

The most striking feature in Figure 6 is that HCRs are mainly distributed along the  
edges of CO(1--0) contours. 
The fact that no HCR is found close to the CO(1--0) peaks is mainly due to 
our selection effect as CO(1--0) peaks often coincide with H \textsc{ii} regions. 
However, the distribution of HCRs is clearly more extended than the CO(1--0) distribution.
The HCR B2--1 is especially interesting. 
Its morphology follows the CO distribution between molecular clouds 
SMC--B2 and LIRS49, but along the periphery of the 
CO peaks.
While no obvious decrease in the stellar distribution is seen in the DSS
R band image, there is a sharp edge of N27 right where B2--1 starts.

The extended HCR distribution around CO(1--0) peaks is reminiscent of the 
$\rm H_2$ distribution derived by Leroy et al. (2007).  
While CO(1--0) and $\rm H_2$ (spatial resolution of 46 pc) 
were found to have similar structure in the SW bar region, the H$_2$ 
distribution was found to be extended by 30 \% in the outer parts of SMC molecular
clouds. One possible reason for this is the selective photodissociation of CO by the
intense radiation field while H$_2$ survives due to its high self-shielding (e.g. Rubio et al. 1993). 
We show the selected HCRs overlaid on the H$_{2}$ surface density map 
of Leroy et al. (2007) in Figure 7. 
The extended HCR distribution is clearly seen in the figure. 
HCRs avoid the H$_{2}$ peaks, which mostly coincide with the CO(1--0) peaks, 
but are still distributed in regions with 
$\rm \Sigma_{H_{2}} = 60 - 180~M_{\odot}~pc^{-2}$.

Bolatto et al. (2005) analyzed multiple CO line ratios of star-forming regions 
in the SMC bar and found that high-temperature molecular gas traced by CO(4--3)
($T_{kin}=100-300$ K and $n_{\rm H_2}=10^{2-3}~\rm cm^{-3}$) 
is frequently associated with colder ($T_{kin}=10-60$ K) but much denser 
($n_{\rm H_{2}}=10^{4-5}~\rm cm^{-3}$) molecular gas.
This suggests that a multi-phase (cold and warm) molecular  
medium exists in this region. 
Therefore, the HCR B2--1 may be tracing a moderately dense and warm envelope 
of the cool SMC--B2 molecular cloud. 

The emerging picture from the above two regions is that HCRs most likely trace 
a moderately dense gas in the outskirts of CO(1--0) clouds, abundant 
in $\rm H_2$, but generally weak in CO(1--0) in regions with strong radiation field, 
and most likely reasonably warm ($T>10$ K).
Since our IR observations have a high angular resolution of 6$''$ (or 3 pc), significantly
higher than what has been achieved for CO or H$_2$, 
HCRs (ranging in size from a few to 50 pc) 
demonstrate that H$_{2}$ gas traced by weak CO emission 
is clumpy on scales of a few pc in the SMC. 

\subsection{HCR LIRS36--EAST}

This particular HCR is the only one visible by eye as a dark silhouette
in the \textit{Spitzer} 24 $\mu$m image and has largely motivated our study. 
The HCR LIRS36--east is located at the east edge of molecular cloud LIRS36
($\alpha$, $\delta$)$_{\rm J2000}=
($$\rm 00^{h}$:$\rm 47^{m}$:$\rm 30^{s}$, 
$\rm -73^{\circ}$:$\rm 07'$:$\rm 30''$) and near SMC--B2. 
Figure 8 shows the HCR LIRS36--east seen as a white, finger-like silhouette 
in the \textit{Spitzer} 24 $\mu$m image. 
White contours are from the NANTEN CO(1--0) Survey (Mizuno et al. 2001). 

In Figure 9 we show $\rm H_{2}$ (28.2 $\mu$m; red), 
S \textsc{iii} (33.5 $\mu$m; blue), and 1.2 cm radio continuum (white) emission 
overlaid on the \textit{Spitzer} 24 $\mu$m image.
Observations of $\rm H_{2}$ and  S \textsc{iii} were obtained with the Infrared
Spectrograph (IRS) on board the \textit{Spitzer} (Bolatto et al., in preparation),
while the radio continuum source at 1.2 cm is a detection of SNR J0047.2--7308
from our study (see Section 5.3).
SNR J0047.2--7308 is visible in 
the radio continuum emission, while SNR B0045--7322 (Ye \& Turtle 1993) is 
traced by S \textsc{iii} north of HCR LIRS36--east.
It is interesting to note the compressed morphology of S \textsc{iii} and $\rm H_{2}$
contours on both north and south side of HCR LIRS36--east. 
This morphology is suggestive of the existence of a dense medium associated 
with HCR LIRS36--east, in agreement with findings for several other HCRs discussed earlier.

In addition, Scalise \& Braz (1982) detected $6_{16}-5_{23}$ transition 
of the water molecule ($\nu = 22.235$ GHz) at a position very close to HCR LIRS36--east 
($\alpha$, $\delta$)$_{\rm J2000}=($$\rm 00^{h}$:$\rm 47^{ m}$:$\rm 31^{s}$, 
$\rm -73^{\circ}$:$\rm 08'$:$\rm 20''$) with the 14-m Itapetinga Radio Telescope (IRT). 
While the existence of water molecule in this region will clearly
signify dense molecular gas, it is possible that
the water maser is associated with one of nearby H \textsc{ii} regions 
since the positional accuracy of the IRT is quite poor (beam size $\sim 3'$).
In addition, Oliveira et al. (2006) performed 22 GHz observations 
toward the pointing of Scalise \& Braz (1982) with the 64-m Parkes telescope
but failed to confirm the water maser detection.  

\section{Search for Molecules in HCR LIRS36--east}

Based on the estimated properties of HCRs (Section 3), 
as well as their comparison with observations at other wavelengths, 
HCRs likely trace reasonably dense molecular gas.
To investigate physical properties of this medium, 
we have undertaken a search for several molecular species in the direction of HCR LIRS36--east. 
This particular HCR was chosen because of its obvious dark silhouette 
in the \textit{Spitzer} 24 $\mu$m image. 
In addition, other multi-wavelength data strongly suggest the existence 
of dense gas in this region. Surprisingly, no spectral lines were detected. 
In this section we summarize our observations and estimate upper limits to the column
densities of several molecular species. 
We did detect a radio continuum source at 1.2 cm, 
corresponding to SNR J0047.2--7308. 
This finding is discussed in Section 5.3.

\subsection{Observations of NH$_{3}$ with the ATCA}

Observations of (1,1) and (2,2) inversion transitions of ammonia ($\rm NH_{3}$)
were made on 2007 September 26--29 
using the Australia Telescope Compact Array (ATCA)\footnote{The Australia 
Telescope Compact Array is part of the Australia Telescope which is funded by the Commonwealth of
Australia for operation as a National Facility managed by CSIRO.}. 
Six (effectively five; the sixth antenna is down-weighted due to its long baseline) 
antennas in the H75 configuration were used during the observations. 
The primary beam of the ATCA in the H75 configuration at 1.2 cm is $2.4'$.
The synthesized beam size is $24.6'' \times 22.4''$. 
Two intermediate frequencies (IFs) were used to measure both transitions simultaneously. 
The first IF ($\Delta \nu=8$ MHz; 102 $\rm km~s^{-1}$) was tuned
to the frequency of 23.694 GHz with 512 channels 
($\Delta \nu=16$ kHz; 0.2 $\rm km~s^{-1}$) to sample  
emission from the (1,1) transition. 
The second IF ($\Delta \nu=16$ MHz; 205 $\rm km~s^{-1}$) was set 
at the rest frequency of 23.723 GHz with 256 channels 
($\Delta \nu=63$ kHz; 0.8 $\rm km~s^{-1}$) to sample the emission 
from the (2,2) transition. 
As shown in Figure 8, HCR LIRS36-east is well enclosed by the ATCA primary beam 
and we have used only a single-pointing for our observations.
The total integration time on the source was 32 hours. 
PKS 2353--686 was observed every 30 minutes to calibrate the phase and gain. 
The flux calibration source was PKS 1934--638. 
Observations of PKS 1921--293 were made to calibrate the bandpass. 
 
The data reduction including flagging, calibration of antenna amplitude and 
phase was made by using the radio interferometry reduction package, 
MIRIAD (Sault et al. 1995). 
We followed the procedure outlined in the MIRIAD manual for 1.2 cm ATCA 
observations\footnote{http://www.atnf.csiro.au/computing/software/miriad/userguide/node186.html}. 
We achieved a sensitivity of $\sim$ 1 mJy per synthesized beam per channel for both transitions. 

\subsubsection{Upper limits to the column density of NH$_{3}$}

We did not detect emission from the either (1,1) or (2,2) transitions. 
Three-channel Hanning smoothing to reduce the noise per spectral resolution element
was also unsuccessful in yielding any detection of emission. 
Table 1 shows 3$\sigma$ noise levels in the non-smoothed image.   

With 3$\sigma$ values, we can set upper limits to the optical depths of the 
(1,1) and (2,2) transitions. 
The upper limits can be determined from the basic radiative transfer equation
\begin{equation}
T_{mb}=T_{bg}e^{-\tau}+T_{ex}(1-e^{-\tau})
\end{equation}

\noindent where $T_{mb}$ is the main beam brightness temperature of each transition, 
$T_{bg}$ is the background temperature, 
$T_{ex}$ is the excitation temperature of the observed transition, 
and $\tau$ is its optical depth. 
Since we have to subtract the continuum emission to be able to detect 
any line emission over the continuum, an observable quantity $T_{mb}^{'}$ can be 
defined as $T_{mb}^{'}=T_{mb}-T_{c}$ where $T_{c}$ is the brightness temperature of 
the continuum emission. Furthermore, we assumed that $T_{c} = T_{cmb} = 2.73$ K. 
Hence, equation (2) can be written as

\begin{equation}
T_{mb}^{'}=(T_{ex}-T_{cmb})(1-e^{-\tau})
\end{equation}  

\noindent To obtain upper limits, we assumed the excitation temperature for the (1,1) 
and (2,2) transitions, $T_{ex}=20$ K (Frieswijk et al. 2007). 
The column densities of the (1,1) and (2,2) transitions were derived by using the following 
equations (Mangum, Wootten, \& Mundy 1992)
\begin{eqnarray}
N(1,1)=6.60\times10^{14}~\frac{T_{ex}(2,2;1,1)}{\nu(1,1)}~\tau(1,1)\Delta v~cm^{-2} \nonumber \\
N(2,2)=3.11\times10^{14}~\frac{T_{ex}(2,2;1,1)}{\nu(2,2)}~\tau(2,2)\Delta v~cm^{-2}
\end{eqnarray} 

\noindent where $\nu(1,1)$ and $\nu(2,2)$ are the transition frequencies in GHz, 
$\tau(1,1)$ and $\tau(2,2)$ are the optical depths of the main hyperfine 
components of the (1,1) and (2,2) transitions, and $\Delta v$ is the velocity 
width in $\rm km~s^{-1}$. From the recent detection of NH$_{3}$ in the LMC
(Henkel \& Ott 2009), we adopted $\Delta v=5~\rm km~s^{-1}$. 
Table 1 shows the derived upper limits to the column densities of the two transitions.

\subsection{Observations of $\rm \bf N_{2}H^{+}$, HNC, $\rm \bf HCO^{+}$, 
and HCN with the Mopra Telescope}

Observations of $\rm N_{2}H^{+}$(1--0), HNC(1--0), $\rm HCO^{+}$(1--0), 
and HCN(1--0) lines were made on 2008 May 5--7 
using the 22-m Mopra single-dish radio telescope\footnote{
The Mopra radio telescope is part of the Australia Telescope which is funded by the Common-
wealth of Australia for operation as a National Facility managed by CSIRO. The University of
New South Wales Digital Filterbank used for the observations with the Mopra Telescope was
provided with support from the Australian Research Council.}. 
All 4 lines are good tracers of dense and cold gas 
($n_{crit} = 10^{5-6}~\rm cm^{-3}$). 
The zoom mode of the Mopra Spectrometer (MOPS) was used for these observations. 
The bandwidth and resolution of the zoom mode at 90 GHz are 456 $\rm km~s^{-1}$ 
and 0.11 $\rm km~s^{-1}$ (137 MHz and 33 kHz) respectively. 
The beam size of the Mopra telescope at 90 GHz is  $\sim 36''$ (Ladd et al. 2005). 
Observations of two positions with the highest contrasts were performed. 
The positions of these two pointings are ($\alpha$, $\delta$)$_{\rm J2000}=
($$\rm 00^{h}$:$\rm 47^{m}$:$\rm 24.6^{s}$, $\rm -73^{\circ}$:$\rm 07'$:$\rm 34''$) 
and ($\rm 00^{h}$:$\rm 47^{m}$:$\rm 30.4^{s}$, $\rm -73^{\circ}$:$\rm 07'$:$\rm 28''$). 
The first and second pointings were integrated for 2.6 hours and 2.4 hours, on source,
respectively. 
The system temperature varied from 170 to 260 K during the observations. 

The initial data reduction of correcting the spectra for the off-source bandpass,
and averaging, was done with the ATNF Spectral line Analysis Package, 
ASAP\footnote{http://www.atnf.csiro.au/computing/software/asap/}.
Further data reduction including summing spectra, fitting and subtracting of 
baseline ripples, and smoothing spectra was done by using generic IDL routines 
and \textsc{mpfitfun} (Markwardt 2009). 
We achieved a sensitivity of $\sim$ 15 mK in the main beam brightness temperature scale 
per channel for 4 lines.  

\subsubsection{Upper limits to the column densities of $N_{2}H^{+}$, HNC, $HCO^{+}$,
and HCN}

We did not detect emission from any of the 4 lines. 
Smoothing spectra with 20 channels also failed to yield emission. 
Table 2 shows 3$\sigma$ upper limits to the column densities of all transitions. 
These values were calculated in the Rayleigh-Jeans and optically thin approximation, 
using the following equation
\begin{equation}
N=\frac{8 \pi W}{\lambda^{3} A} \times \frac{g_{l}}{g_{u}} \times 
\frac{1}{J_{\nu}(T_{ex})-J_{\nu}(T_{bg})} \times 
\frac{1}{1-exp(-h \nu /k T_{ex})} \times \frac{Q_{rot}}{g_{l}~exp(-E_{l}/k T_{ex})}
\end{equation}

\noindent where $W$ is the integrated intensity of the line 
($W = \pi^{1/2}\Delta v T_{mb}/2\sqrt{ln2}$ for a Gaussian line), 
$\lambda$ is the rest wavelength of the transition, 
$A$ is the Einstein coefficient, $g_{l}$ and $g_{u}$ are the statistical weights of the 
lower and upper levels, $J_{\nu}(T_{ex})$ and $J_{\nu}(T_{bg})$ are the equivalent
 Rayleigh-Jeans excitation and background temperatures, 
 $Q_{rot}$ is the partition function, 
 and $E_{l}$ is the energy of the lower level from the ground rotational level. 

Throughout our calculation, we adopted $A$ and $Q_{rot}$ values from 
molecular spectroscopy data of the Jet Propulsion Laboratory\footnote{http://spec.jpl.nasa.gov/}. 
$T_{ex}=20$ K and $T_{bg}=2.73$ K were assumed for the calculation. 
$\Delta v=4$ $\rm km~s^{-1}$ was adopted based on observations by Chin et
al. (1998) of the nearby molecular cloud LIRS36. 
Table 2 shows the calculated upper limits to the column densities. 

\subsection{Detection of Continuum Emission from SNR J0047.2--7308}

The only detection in our observations is a radio continuum source at 1.2 cm. 
The location of the source coincides with the well-known supernova remnant in
the SMC, SNR J0047.2--7308 (Dickel et al. 2001), centered at 
($\alpha$, $\delta$)$_{\rm J2000}=
($$\rm 00^{h}$:$\rm 47^{m}$:$\rm 17^{s}$, $\rm -73^{\circ}$:$\rm 08'$:$\rm 43''$). 
While a wealth of data at other wavelengths exists for this SNR, 
this is the first detection at 1.2 cm.
The 1.2 cm continuum image shown in Figure 9 was obtained by summing over all channels 
in the two ATCA data cubes for the (1,1) and (2,2) transitions. 
The estimated flux density of the SNR at 1.2 cm is $7.4 \pm 0.5$ mJy. 

SNR J0047.2--7308 has been observed at several different radio frequencies 
(408 MHz: Clarke et al. 1976; 843 MHz: Ye 1988; 
1 GHz, 1.34 GHz, and 2.4 GHz: Dickel et al. 2001; 
1.42 GHz, 2.37 GHz, 4.8 GHz, and 8.64 GHz: Payne et al. 2004). 
Figure 10 shows 23 GHz (1.2 cm) and 4.8 GHz radio continuum emission, 
in white and thick-grey contours respectively, overlaid on 
the 1.34 GHz image. The 1.2 cm continuum source coincides with most of 
the northern rim of the SNR. On the NE side of the rim, the 1.2 cm
contours follow well the morphology of the rim, however the peak of the
1.2 cm emission is off-set slightly ($\sim30''$) from the peak of the northern rim.

Figure 11 shows the radio spectrum of the SNR.
Error bars show available flux measurements from the literature
at several frequencies.
Our flux density measurements at 1.34 GHz, 4.8 GHz, and 23 GHz,
obtained only for the area enclosed by the 23 GHz radio emission (which covers 
most of the northern rim of the SNR), are shown as circles in the same figure.
These measurements are very similar to the values from the literature derived
for the extent of the whole SNR, suggesting that most of the SNR flux is 
in its northern rim, as also evident from Figure 10.
The solid line in Figure 11 corresponds to the spectral index of $-0.6\pm 0.2$ 
and was derived using the measurements from the literature.
This spectral index agrees with Payne et al. (2004), and falls
within the range expected for SNRs ($-0.8 < \alpha < -0.2$, McGee \& Newton 1972). 

Curiously, our flux density at 23 GHz is at least a factor of 5 lower than what would be 
expected for the SNR based on its spectral index. 
As the largest angular scale we are sensitive to in our observations corresponds to 1$'$ 
and is smaller than the SNR itself (diameter $\sim 2'$), 
it is possible that some flux is missing in our observations, 
resulting in the underestimated flux density at 23 GHz. 
In addition, the SNR is located at the south edge of our field-of-view where
we are 5 times less sensitive than in the image center.
We therefore conclude that both missing short-spacings and low sensitivity
are the most likely reasons for this large departure of the radio spectrum at 23 GHz.
An alternative/additional source for this departure could be
a spatial variation of the spectral index across the SNR.
Dickel et al. (2001) used images at 1.34 and 2.4 GHz to derive the spectral index
image of the SNR. Across the SNR, a significant variation can be seen in their Figure 4, 
with the spectral index steepening from $\alpha=-0.4$ close to the center of the SNR, 
to $\alpha \sim -1$ in several small knots in the northern rim.  

\section{Discussion}

\subsection{Comparison with Other Studies of Dark Clouds in the SMC}

Dobashi et al. (2009) recently constructed an extinction map of the SMC 
using the color excess at near-IR wavelengths ($E(J - H)$). 
They selected ten dark clouds from the $E(J - H)$ map 
(6 in the southern bar, 2 in the eastern wing, and 2 in a region 
just below the northern bar).
Similarly to our results, their distribution of dark clouds is similar to 
that of CO clouds, based on the NANTEN CO(1--0) map. 
However, their dark clouds, in particular those in the southern bar, 
frequently coincide with CO peaks, whereas our HCRs are distributed 
along the peripheries of CO peaks. 
This difference mainly arises from the fact that our method is blind to 
high-extinction regions within H \textsc{ii} regions, which mostly correspond 
to strong CO clouds (see Section 2.3). 
While the two methods are complementary, we emphasize that our spatial resolution 
is much higher; angular resolution of the $E(J - H)$ map is $2.6'$ 
and thus even the largest HCRs in our study 
($\sim 50$ pc or $\sim 2.9'$) can hardly be identified.

Dobashi et al. (2009) also provides independent evidence 
for the nature of the largest HCR found in the N83/N84 region 
(labeled as 1 in the lower panel of Figure 5; HCR N83/N84--1). 
This HCR corresponds to their Cloud I. 
It has $E(J - H) = 0.23 \sim 0.29$ mag or $A_{V} = 2.51 \sim 3.16$ mag. 
This is 2 times higher than our estimate.  
The most likely cause for this discrepancy is the uncertainty 
in the position of the HCR/Cloud I along the line-of-sight.
As we discussed in Section 3, 
we have assumed that all HCRs are located in the middle of the SMC,
while Dobashi et al. (2009) inferred this quantity from a comparison 
of the $E(J - H)$ map with a simple model.

\subsection{Abundances of Molecular Species}

As shown in Table 1, our upper limit to N($\rm NH_{3}$) of HCR LIRS36--east is 
about $6\times10^{12}$ cm$^{-2}$.
This column density is similar to the only detection of $\rm NH_{3}$ 
in the LMC, accomplished very recently with the
upgraded ATCA (Henkel \& Ott 2009) in the direction of N159--W, 
one of the most active star-forming regions in the LMC.
They found two velocity components in the spectrum and 
obtained N($\rm NH_{3}$) of $(4.7\pm 0.5)\times 10^{12}~\rm cm^{-2}$ 
and $(1.1\pm 0.3)\times 10^{12}~\rm cm^{-2}$, 
with rotational temperatures of $17\pm2$ K and $15\pm5$ K, respectively.
Obviously, our ATCA observations achieved a sensitivity comparable to the 
NH$_{3}$ detection in the LMC.
However, considering that the SMC has metallicity a few times lower than the LMC,
its molecular abundance is expected to be lower in general (e.g. Israel et al. 1993). 
Future, more sensitive observations are required to probe and constrain the NH$_{3}$
abundance in the SMC.
Up to present, NH$_{3}$ has not been detected in the SMC. 
 
Our upper limit to N($\rm NH_{3}$) could be significantly underestimated due to the beam dilution 
if dense molecular clouds are significantly smaller than 
the ATCA synthesized beam, an effect first pointed out by Osterberg et al. (1997) 
in their unsuccessful search for NH$_{3}$ in the Magellanic Clouds.
The ATCA beam at 1.2 cm ($\rm synthesized~beam~size \sim 24''$) corresponds to 7 pc 
at the distance of the SMC, but dense molecular clumps are likely to 
be much smaller than the beam.  If this is the case, our upper limit to N($\rm NH_{3}$) 
may be significantly underestimated.
If we assume that molecular clumps within our ATCA beam have a size
of 1 pc (for example, a dense clump $\sim2$ pc in size was observed in the LMC 
by Wong et al. 2006), then the upper limit to N($\rm NH_{3}$) increases to 
a few times $10^{14}~\rm cm^{-2}$.


Our upper limits for other molecular transitions are shown in Table 2.
For $\rm HCO^{+}$ we have achieved an excellent sensitivity. 
The upper limit to N($\rm HCO^{+}$) is $2-6$ times lower than 
previous measurements obtained in the SMC (Chin et al. 1998; Heikkil\"a et al. 1999)
and at least 10 times lower than any detection of the same molecule in the LMC.
Several of earlier SMC and LMC measurements are summarized in Table 3 for 
a comparison with our results. 
Similarly to N(NH$_{3}$), the upper limit to $\rm N(HCO^{+})$ is about 1000 times 
lower than what is found typically for Galactic IRDCs, but is 
similar to $\rm N(HCO^{+})$ for Galactic diffuse clouds.
If we correct for the beam dilution effect (assuming again a cloud size of 1 pc), 
the upper limit to N($\rm HCO^{+}$) reaches a few times $10^{13}$ cm$^{-2}$.

Our upper limits for HCN and HNC are significantly lower than any
detections of these molecules in the LMC, but worse (higher)
than the two previous detections in the SMC  (Chin et al. 1998; Heikkil\"a et al. 1999).
$\rm N_{2}H^{+}$ was never 
detected in the SMC and the only detection of this molecule in the LMC 
has a column density that is 3 times lower than our upper limit. 
Our upper limit to N(HNC) is 3 times higher than the SMC
detection in the quiescent molecular cloud LIRS36, but $4-6$ times lower than several 
detections in the LMC. For HCN, the existent SMC detections show large variations 
in column density, from $1.4\times10^{12}$ cm$^{-2}$ in the star-forming region N27 
to $2\times10^{10}$ cm$^{-2}$ in LIRS36.  
Our upper limit, $< 10^{12}$ cm$^{-2}$, is 
significantly lower than any of the LMC detections.
Clearly, our upper limits for HNC and HCN are lower than the 
column densities detected in N27, 
but higher than detections in LIRS36 and require future, more sensitive observations.

Our upper limits for HCN, HNC and HCO$^+$, which are better than typical LMC detections, 
confirm that molecular abundances in the SMC are generally lower than in the LMC.
In addition, the abundance in both the LMC and SMC can vary spatially by a factor of 
$10 - 100$. 
For example, the molecular abundance of the quiescent molecular cloud LIRS36 is 
significantly lower than the abundance of the star-forming molecular cloud N27.
This effect has been reported previously (Heikkil\"a et al. 1999) and was 
considered to be due to either metallicity variation and/or ISRF variation.
However, it is important to emphasize that molecular multi-line studies
of the Magellanic Clouds are still rare and only a handful of measurements exists.

Four molecular species in our study (all except HCO$^{+}$) contain nitrogen (N). 
NH$_3$ and N$_2$H$^+$ were never detected in the SMC, while HCN and HNC
were detected only in two molecular clouds.
Several authors have noticed a low abundance of N-containing species
in the SMC and LMC and have suspected that this is due to a very small
nitrogen abundance in these galaxies (e.g. Chin et al. 1998; Wang et al. 2009).
It was not clear though whether nitrogen abundance is exceptionally low
in the handful of observed molecular clouds, or this is due to 
chemical processes which operate in all molecular clouds. 
Future, deeper observations will be required to constrain molecular abundance 
in HCR LIRS36--east,
however based on our current non-detections its conditions are more similar, or even more
extreme, to those of the quiescent molecular cloud LIRS36 than the star-forming cloud N27. 
The exceptionally low abundance of nitrogen makes the detection of any N-containing 
molecular species in the SMC extremely hard. 

\subsection{Constrainst on the Physical Properties of HCRs}

What can we learn from our non-detections?
Observed molecular species probe gas with a range of physical properties.
CO is considered as a good tracer for low-density molecular gas with 
a molecular hydrogen density $n\sim10^{3}$ cm$^{-3}$, while NH$_{3}$ is a good tracer for 
moderately dense gas with a density of $10^{3-4}$ cm$^{-3}$.
N$_{2}$H$^{+}$, HNC, HCO$^{+}$, and HCN trace 
dense and cold gas with $n \sim 10^{5-6}$ cm$^{-3}$.
Our strong upper limit to N(HCO$^+$), as well as the 
presence of weak CO(1--0) emission, suggest that the density 
in HCR LIRS36--east is likely to be $n \sim 10^{3}$ cm$^{-3}$. 
This agrees with the morphological hints in Section 4.
Hence, at least one of our HCRs is most likely moderately dense with
$n\sim10^{3}$ cm$^{-3}$.
This also agrees with the estimated $\rm H~\textsc{i} + H_{2}$ density of 
$(0.5 - 2) \times 10^{3}~\rm cm^{-3}$
based on the H$_2$ image of Leroy et al. (2007), the 
H \textsc{i} surface density map of Stanimirovi\'c et al. (1999), and assuming
a cloud size of $\sim$ 10 pc.

The hydrogen volume density ($\sim 10^{3}~\rm cm^{-3}$) of HCRs agrees with the expectation for 
the cool atomic medium in a low metallicity environment.
For example, Wolfire et al. (1995) estimated, based on the heating and cooling equilibrium,
that for a dust-to-gas ratio $\sim30$ times lower than in the Galaxy, 
the cold neutral medium (CNM) and the warm neutral medium (WNM) can co-exist only at a 
significantly high-pressure ($\sim10^4$ K cm$^{-3}$). This results
in the CNM temperature being as low as 25 K and a volume density
being as high as $1.2\times10^3$ cm$^{-3}$. 
Dickey et al. (2000) measured the CNM temperature of 40 K or less in the SMC, 
in agreement with the Wolfire et al. (1995) prediction. 
Much of the cold gas in the SMC comes from regions deep inside interstellar clouds, 
at cold temperatures but high-densities, which in the Galaxy would be totally 
dominated by H$_2$. 

In addition, HCRs appear to have roughly an equal column density of 
H $\textsc{i}$ and $\rm H_{2}$, a few times $10^{21}$ cm$^{-2}$
(see Section 3). 
This is consistent with what Dickey et al. (2000) predicted for 
the transition 
between H \textsc{i}-dominated and molecule-dominated 
clouds 
in the SMC. In the Galaxy, molecular clouds achieve 
enough self-shielding from the interstellar radiation
field at a much lower hydrogen column density, $\sim 4 \times 10^{20}$ cm$^{-2}$ 
(see Draine \& Bertoldi 1996 and references therein).
As atomic clouds are converted into molecular clouds, 
which consequently become fuel for star formation, 
properties of such transition clouds are
important for constraining the process of gas condensation and collapse.
We suggest that HCRs are good candidates for 
future observations with the \textit{Herschel Space Observatory}
of species such as neutral carbon (C), which are good tracers of translucent gas.

\subsection{Are HCRs Counterparts of Galactic IRDCs?}

As pointed out in Section 6.1, our non-detections of various
molecular species could be due to significant clumping of molecular gas inside 
HCRs. In addition, if HCRs are located at the back of the SMC, our extinction
values are significantly under-estimated. Therefore, we cannot exclude
the possibility of HCRs being counterparts of IRDCs found in the Galaxy.

What are the expected molecular abundances for dark clouds in the SMC?
The only chemical model that investigated molecular abundances of dark clouds (with $A_V\sim10$ mag) 
in the SMC is by Millar \& Herbst (1990). This work showed that the abundance of various 
molecular species does not simply scale with the metal abundance.
While the abundance of the N-containing species is often more directly related to the total nitrogen 
abundance, the abundance of hydrocarbons (like HCO$^+$) has a more complex dependance on
metallicity. In steady state, the abundance of NH$_{3}$ is expected
to scale roughly with the abundance of nitrogen, while
the abundance of HCO$^+$ is expected to be roughly similar to that in dark clouds
in the Galaxy. While the direct model predictions depend heavily
on the exact input parameters, we expect that these trends will hold as direct
consequences of interstellar chemical reactions.
We hence expect the abundance of NH$_{3}$ to be at least 10 times lower in the
SMC than in the Galaxy (based on the nitrogen abundance from 
Millar \& Herbst 1990), while the abundance of HCO$^+$ should be roughly 
similar in the SMC and the Galaxy.

Galactic IRDCs have N($\rm NH_{3}) \sim10^{15}~cm^{-2}$ (Pillai et al. 2006; Frieswijk et al. 2007)
and N($\rm HCO^{+}$) $\sim$ a few times $10^{14}$ cm$^{-2}$ (Purcell et al. 2006).
In the case of the SMC, after accounting for the beam dilution effect 
(due to molecular clumps $\sim1$ pc in size), we arrived at the upper limits on 
N($\rm NH_{3}) \sim$ a few times $10^{14}~\rm cm^{-2}$ and
N($\rm HCO^{+}) \sim$ a few times $10^{13}~\rm cm^{-2}$. 
Our upper limit on N(HCO$^+$) is especially stringent.
While our upper limit on the abundance of NH$_{3}$ in HCRs is in agreement
with what would be expected for IRDCs in the SMC (after accounting for metallicity), 
the abundance of HCO$^+$ is at least a factor of 10 lower.
We conclude that HCRs are not counterparts of Galactic IRDCs, unless 
HCO$^+$ emitting gas contains tiny clumps,  $\sim0.5$ pc or smaller in size, and/or
there are significant spatial variations in metallicity across the SMC.
Only future high-resolution observations 
(e.g. with the Atacama Large Millimeter Array) 
will be able to investigate this possibility.

\subsection{Alternative Explanation for HCRs} 


An alternative possibility which may be able to explain properties of HCRs
involves an anomalous dust abundance yielding an unusually low intensity of
24 $\mu$m emission.
This would happen if for example there are significant spatial variations
in the abundance of various dust species. 
The 24 $\mu$m emission in both the SMC and the LMC is dominated by emission from VSGs.
In the LMC, the recent \textit{Spitzer} observations have shown 
regions of 70 $\mu$m excess relative to the 24 $\mu$m emission (Bernard et al. 2008). 
The most likely interpretation of this excess is an increased abundance of 
large VSGs relative to smaller VSGs, which could be caused by erosion of even larger grains by shocks. 
This illustrates just one of several possible physical processes that could result 
in an anomalous VSG abundance. 
Interestingly, the 70 $\mu$m excess sources in the LMC do not correlate well 
with molecular clouds. 
This is different from HCRs in the SMC where the largest number of objects is found
around CO clouds. 

To examine the possible anomalous dust abundance in HCRs, 
we measured flux densities at IR wavelengths provided by  
the \textit{IRAS} images (Stanimirovi\'c et al. 2000) 
and the \textit{Spitzer} images (Bolatto et al. 2007)
for several selected HCRs, including HCR LIRS36--east, HCR B2--1, and HCR N83/N84--1. 
We examined various flux density ratios; an anusually low value of the 
24--to--160 $\mu$m flux density ratio would be an especially good indicator
for the deficiency of VSGs relative to the large dust grains.
However, we found that the 24--to--160 $\mu$m flux density ratio
of several HCRs is similar to the average SMC ratio,
suggesting that VSGs are not unusually deficient in HCRs. 
It is important to emphasize that more detailed measurements of flux 
ratios are hampered by coarse resolution of IR images at long wavelengths (100 and 160 $\mu$m)
relative to the 24 $\mu$m image. In addition, 
the proximity of HCRs to bright H \textsc{ii} regions makes
the background subtraction highly uncertain. 
Further analysis of the SED of HCRs at long IR wavelengths with higher-angular resolution
will be possible with the \textit{Herschel Space Observatory} 
and would allow us to investigate further this possibility.

\section{Summary and Conclusions}

We have searched for dense ISM in the SMC by applying 
the unsharp-masking technique to the \textit{Spitzer} 24 $\mu$m image. 
When applied to Galactic IR images, the unsharp-masking technique  
reveals the highest extinction regions, IRDCs, which are
very dense ($n > 10^{5}~\rm cm^{-3}$) and cold ($T < 25$ K) interstellar clouds.  
The unsharp-masking technique is limited to regions with 
diffuse IR background and is blind to dense gas within boundaries of H \textsc{ii} regions.
As H \textsc{ii} regions are often associated with CO peaks, we are not sensitive
to molecular gas with substantial CO emission but are mainly tracing a population 
of interstellar clouds located in the outskirst of CO/H \textsc{ii} regions.
We call these clouds  high-contrast regions (HCRs)
and our main results are as follows. 

1. We have identified a population of 55 HCRs 
from the decremental contrast image.  
HCRs have a highly inhomogeneous spatial distribution, 
39 are located in the southern bar, 10 in the northern bar, and
6 in the eastern wing.  This distribution is qualitatively similar to the distribution
of CO clouds and is suggestive of a possible physical connection between
the two populations of clouds. 
While HCRs with low contrast values may be spourious detections,
this correlation with CO clouds gives us confidence that
most HCRs are distinct physical entities. 

2. Most identified HCRs have a size of $3 - 18$ pc, which is similar to  
typical or large Galactic IRDCs. 
The distribution of peak-contrasts (at 24 $\mu$m) of HCRs has a peak at $2 - 2.5$ \% and 
this is a factor of $2 - 3$ lower than that of Galactic IRDCs,
suggesting only moderately dense environments ($A_V\sim1$ mag).
HCRs in the SMC are less dark than Galactic IRDCs.

3. In the southern bar region, HCRs are mainly distributed in the outskirts of CO clouds
and have integrated CO(1--0) intensity of $0.1 - 0.5$ K km s$^{-1}$. 
The H$_{2}$ surface density map (Leroy et al. 2007), however, 
suggests they are abundant in H$_{2}$  ($\rm \Sigma_{H_{2}} = 60 - 180~M_{\odot}~pc^{-2}$).

4. To constrain the physical properties of HCRs, we have performed NH$_{3}$ 
(with the ATCA), N$_{2}$H$^{+}$, HNC, HCO$^{+}$, and HCN 
observations (with the Mopra telescope) toward one of HCRs, HCR LIRS36--east. 
While we did not detect any molecular emission, 
we achieved excellent sensitivities 
for NH$_{3}$ and HCO$^{+}$.  
Our upper limits to the column density of various molecular species 
highlight the need for future, more sensitive molecular line observations,
especially for N-containing molecules.
Our upper limits also place a constraint on the density of HCRs, 
$n \sim 10^{3}~\rm cm^{-3}$, suggesting that they are lower density 
environments than Galactic IRDCs.

5. 
HCRs appear to trace regions where about half the hydrogen is molecular.
Due to the low metal abundance, typical CNM 
in the SMC (usually traced through 21-cm absorption lines) has temperature similar to 
what is found for Galactic molecular clouds but much higher density 
as the multi-phase medium can exist only at a higher pressure (Wolfire et al. 1995). 
The estimated volume density ($\sim 10^{3}~\rm cm^{-3}$) 
of HCRs is in agreement with the theoretical expectation for the CNM 
in an environment with a dust-to-gas ratio about 30 times lower than in the Galaxy.
The H \textsc{i} column density, a few times $10^{21}$ cm$^{-2}$, 
suggests that HCRs trace clouds where H$_{2}$ is enough shielded 
against photodissociation. As such clouds
are an essential stage for star formation, searching for HCRs in IR images 
may be a significant way to study early phases of gas condensation 
in low metallicity galaxies.

6. Two alternative explanations may account for the observed properties of HCRs.
Firstly, HCRs could be counterparts of Galactic IRDCs consisting of 
very small molecular clumps (size $\sim0.5$ pc).
If only a tiny portion of the cloud surface area produces emission, 
severe beam dilution may be causing our non-detections of several molecular species.
Secondly, an unusual abundance of dust grains, may result in the absence of
emission at 24 $\mu$m and appearance of HCRs.

7. We have detected a radio continuum source at 1.2 cm, 
at a position that corresponds to the northern portion of SNR J0047.2--7308. 
The measured flux density of the continuum source is $7.4 \pm 0.5$ mJy. 
Using the available literature data for frequencies below 10 GHz, 
we estimated the spectral index of $-0.6\pm0.2$. 
Based on this estimation, the flux density at 23 GHz is at least a factor of 5 lower 
than what would be expected from an extrapolation of the low frequency spectral 
index to 1.2 cm. 
Some missing short-spacings, and poor sensitivity at the position of the SNR, 
and the possible spatial variation of the spectral index across the SNR 
can be responsible for the underestimated flux density at 23 GHz. 

\acknowledgments
This work is based (in part) on observations made with the \textit{Spitzer Space Telescope}, 
which is operated by the Jet Propulsion Laboratory, California Institute of 
Technology under NASA contract 1407. 
Support for this work was provided by NASA through contract 1289519 issued by JPL/Caltech. 
We thank the anonymous referee for suggestions that improved this work. 
We also wish to thank Jay Gallagher, Tony Wong, Monica Rubio, 
Francois Boulanger for useful discussions, 
John Dickel for his kind offer of multi-frequency radio data of SNR J0047.2--7308, 
and Adam Leroy for his 
gracious offer of the SMC H$_{2}$ surface density map and  
comments that improved the manuscript.

\clearpage

\begin{table}
\begin{center}
{\bf TABLE 1} \\
{\small \textsc{Summary of ATCA Observations}} \\
\vskip 0.2cm
\begin{tabular}{c c c c c}\hline \hline
{\footnotesize \textsc{Molecule}} & {\footnotesize \textsc{Transition}} &
{\footnotesize \textsc{Rest Frequency }} & {\footnotesize 3$\rm \sigma$
  $\rm \textsc{Level}^{1}$} & {\footnotesize
  \textsc{ NH$_{3}$ Column Density}$^{2}$} \\ 
 &  & ({\footnotesize MHz}) &  ({\footnotesize $\rm mJy~Beam^{-1}$}) & 
 ({\footnotesize $\rm cm^{-2}$}) \\ \hline 
$\rm NH_{3}$ & (1,1) & 23694.506 & 3.2 & $< \rm 2.7 \times 10^{12}$ \\
$\rm NH_{3}$ & (2,2) & 23722.634 & 1.7 & $< \rm 6.5 \times 10^{11}$ \\
\hline
\end{tabular}
\end{center}
{\footnotesize $\rm {}^{1}$ Noise levels at central channels for frequency
resolutions of 16 kHz and 63 kHz, respectively for (1,1) and (2,2) transitions.} \\
{\footnotesize $\rm {}^{2}$ When (A15) of Ungerechts et al. (1986) is used, 
total ammonia column density, N(NH$_{3}$) is estimated as 
$6.2 \times 10^{12}~\rm cm^{-2}$.}
 \end{table}
 
\begin{table}
\begin{center}
{\bf TABLE 2} \\
{\small \textsc{Summary of Mopra Observations}}\\ 
 \vskip 0.2cm
 \begin{tabular}{
 l 
 c 
 c 
 c
 c 
 c 
 c
 c 
 c 
}\hline \hline
{\footnotesize \textsc{Molecule}} & {\footnotesize \textsc{Transition}} & {\footnotesize \textsc{Rest Frequency (MHz)}} & & \multicolumn{2}{c}{\footnotesize 3$\rm \sigma$ $\rm \textsc{Level}^{1}$ (mK)} & & \multicolumn{2}{c}{\footnotesize \textsc{Column Density} ($\rm cm^{-2}$)}\\ 
\cline{5-6}
\cline{8-9} \\
 & & & & {\footnotesize \textsc{Pointing 1$^{2}$}} & 
{\footnotesize \textsc{Pointing 2$^{3}$}} & & {\footnotesize \textsc{Pointing 1}} & {\footnotesize \textsc{Pointing 2}} \\
 \hline 
$\rm N_{2}H^{+}$ & J=1--0 & 93173.809 & & 47 & 42 & & $< \rm 2.7 \times 10^{12}$ & $< \rm 2.4 \times 10^{12}$\\
HNC & J=1--0 & 90663.574 & & 35 & 31 & & $< \rm 2.9 \times 10^{11}$ & $< \rm 2.5 \times 10^{11}$\\
$\rm HCO^{+}$ & J=1--0 & 89188.518 & & 48 & 47 & & $< \rm 3.5 \times 10^{11}$ & $< \rm 3.4 \times 10^{11}$\\
HCN & J=1--0 & 88631.847 & & 45 & 36 & & $< \rm 1.2 \times 10^{12}$ & $< \rm 9.6 \times 10^{11}$\\
\hline
\end{tabular}
\end{center}
{\footnotesize $\rm {}^{1}$ Noise levels for an individual channel in the main beam brightness temperature scale.} \\
{\footnotesize $\rm {}^{2}$ The position is ($\alpha$, $\delta$)$_{\rm J2000}=
($$\rm 00^{h}$:$\rm 47^{m}$:$\rm 24.6^{s}$, $\rm -73^{\circ}$:$\rm 07'$:$\rm 34''$).} \\
{\footnotesize $\rm {}^{3}$ The position is ($\alpha$, $\delta$)$_{\rm J2000}= 
($$\rm 00^{h}$:$\rm 47^{m}$:$\rm 30.4^{s}$, $\rm -73^{\circ}$:$\rm 07'$:$\rm 28''$).} \\
\end{table}

\begin{sidewaystable}
\begin{center}
{\bf Table 3} \\
{\small \textsc{$\rm N_{2}H^{+}$, HNC, $\rm HCO^{+}$, and HCN Column Densities (Comparison with Other Magellanic Clouds Studies)}}
\vskip 0.2cm
\begin{tabular}{
l 
c
c 
c
>{\centering}p{17mm} 
>{\centering}p{17mm} 
>{\centering}p{17mm} 
c
>{\centering}p{19mm} 
>{\centering}p{17mm} 
} \hline \hline
{\footnotesize \textsc{Quantities}} & & 
{\footnotesize \textsc{Our Work}} & & 
\multicolumn{3}{c}{\footnotesize \textsc{LMC}} 
& &  
\multicolumn{2}{c}{\footnotesize \textsc{SMC}} \\ \cline{5-7} \cline{9-10} \\
& & & & 
{\footnotesize \textsc{N159--W$^{\rm 1}$}} &
{\footnotesize \textsc{30Dor--10$^{\rm 1}$}} &
{\footnotesize \textsc{N113$^{\rm 2}$}} 
& &
{\footnotesize \textsc{N27$^{\rm 1}$}} &
{\footnotesize \textsc{LIRS36$^{\rm 3}$}}
\tabularnewline \hline
N($\rm N_{2}H^{+}$) ($\rm cm^{-2}$) & & $< 2 \times 10^{12}$ & & - & - & $6.3 \times 10^{11}$ & & - & - \tabularnewline
N($\rm HNC$) ($\rm cm^{-2}$) & & $< 3 \times 10^{11}$ & & $4.7 \times 10^{12}$ & $1.3 \times 10^{12}$ & $2.5 \times 10^{12}$ & & $< 3.7 \times 10^{11}$ & $9 \times 10^{10}$ \tabularnewline
N($\rm HCO^{+}$) ($\rm cm^{-2}$) & & $< 3 \times 10^{11}$ & & $1.4 \times 10^{13}$ & $6.6 \times 10^{12}$ & $4.0 \times 10^{12}$ & & $1.9 \times 10^{12}$ & $7 \times 10^{11}$ \tabularnewline
N(HCN) ($\rm cm^{-2}$) & & $< 10^{12}$ & & $1.2 \times 10^{13}$ & $4.8 \times 10^{12}$ & $6.3 \times 10^{12}$ & & $1.4 \times 10^{12}$ & $2 \times 10^{10}$ \tabularnewline
\hline
\end{tabular}
\end{center}
{\footnotesize $\rm {}^{1}$ Heikkil\"a et al. (1999); Column densities are obtained from LTE analysis.} \\
{\footnotesize $\rm {}^{2}$ Wang et al. (2008).} \\
{\footnotesize $\rm {}^{3}$ Chin et al. (1999); Column densities are obtained from LTE analysis. $T_{kin}=T_{ex}=20$ K is assumed.} \\
\end{sidewaystable}
\clearpage


\begin{figure*}
\begin{center}
\includegraphics[scale=.55]{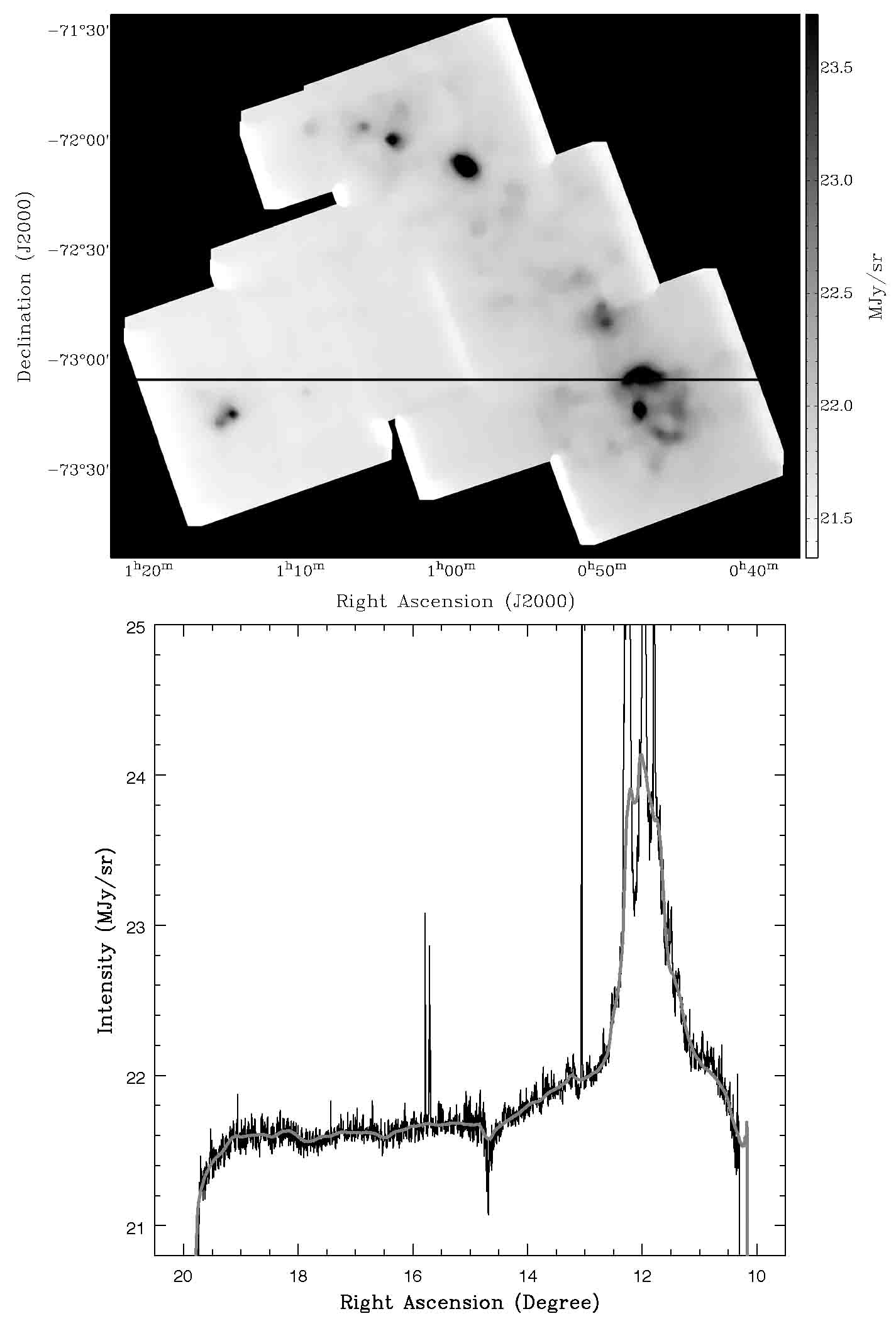}
\vskip 0.3 cm
\begin{minipage}{16cm}
\begin{footnotesize}
\textsc{Fig. 1---} (\textit{Top}) The IR background obtained using the 
median filtering with 63 pixels. 
The horizontal line shows the location of the profile shown in the bottom panel. 
(\textit{Bottom}) The profile of the IR background (grey) 
superposed on the original \textit{Spitzer} image data (black).
Foreground emission (a sum of zodiacal light, Galactic dust emission, and small cosmic 
IR background, $\sim 21.6$ MJy/sr) is not subtracted.  
The gap at $\rm RA = 15^{\circ}$ is due to a small gap between different mosaic
pieces of the \textit{Spitzer} 24 $\mu$m image. 
\end{footnotesize}
\end{minipage}
\end{center}
\end{figure*} 

\begin{sidewaysfigure}
\begin{center}
\includegraphics[scale=.42]{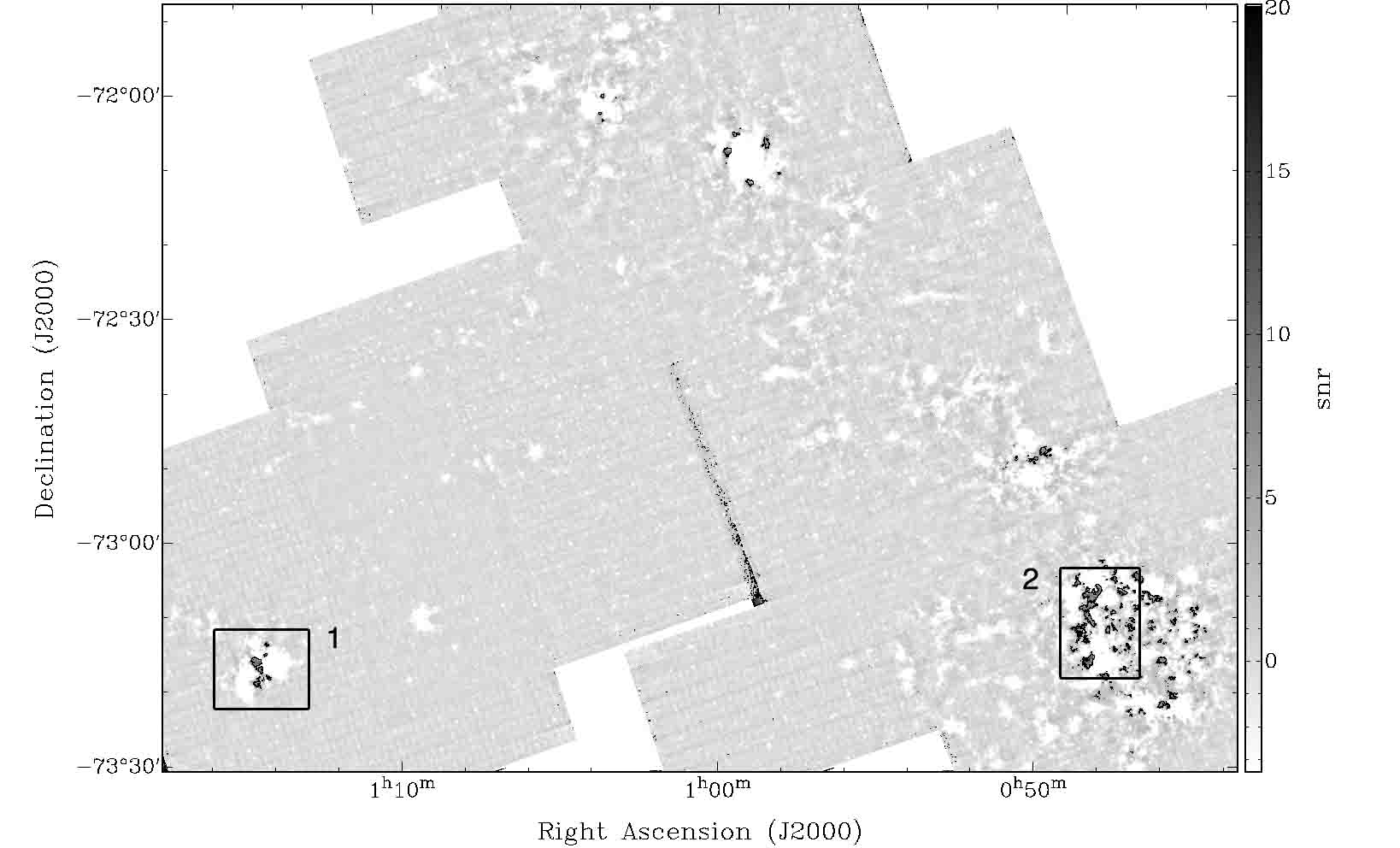}
\vskip 0.3 cm
\begin{minipage}{20cm}
\begin{footnotesize}
\textsc{Fig. 2---} The signal-to-noise ratio image derived from the \textit{Spitzer} 24
$\mu$m image. The black contours are at 5$\sigma$ level. 
H \textsc{ii} regions and stars have values higher than the IR background and 
appear as negative (white) pixels in this image. 
Regions with intensity lower than the IR background (HCRs) appear 
as positive (dark) features in this image. 
Boxes 1 and 2 show selected sub-regions for Figures 5 and 
6 (Box 1: Figure 5; Box 2: Figure 6). 
\end{footnotesize}
\end{minipage}
\end{center}
\end{sidewaysfigure}

\begin{figure*}
\begin{center}
\includegraphics[scale=.5]{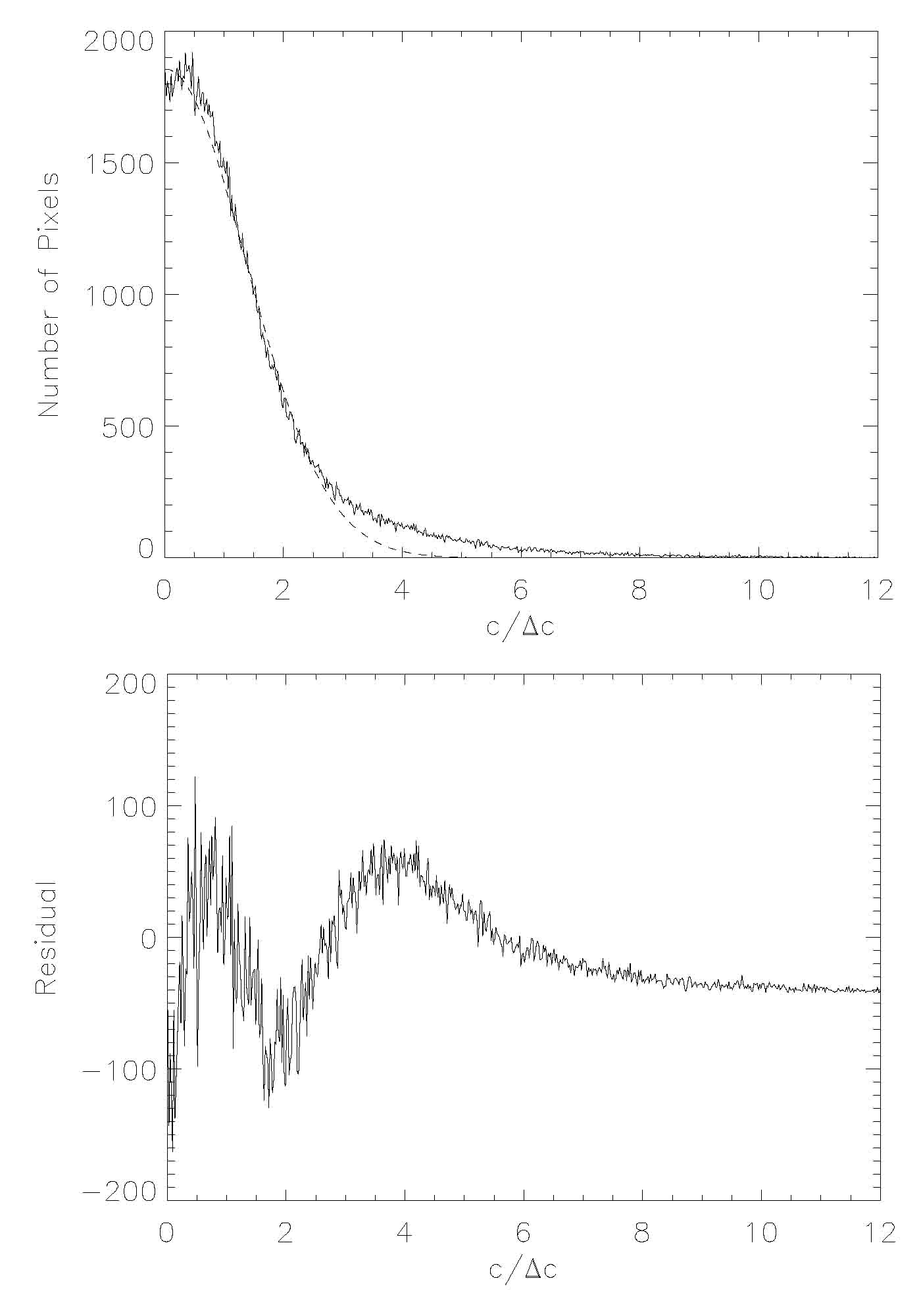}
\vskip 0.3 cm 
\begin{minipage}{16cm}
\begin{footnotesize}
\textsc{Fig. 3---} (\textit{Top}) The histogram of the signal-to-noise ratio image
for the SMC SW bar region. 
The fitted Gaussian function is shown with the dashed line. 
(\textit{Bottom}) The residuals after removing the Gaussian fit. 
From about 3$\sigma$ there is a certain fraction of pixels with significant deviation 
from the Gaussian statistics. 
\end{footnotesize}
\end{minipage}
\end{center}
\end{figure*}

\begin{figure*}
\begin{center}
\includegraphics[scale=0.5]{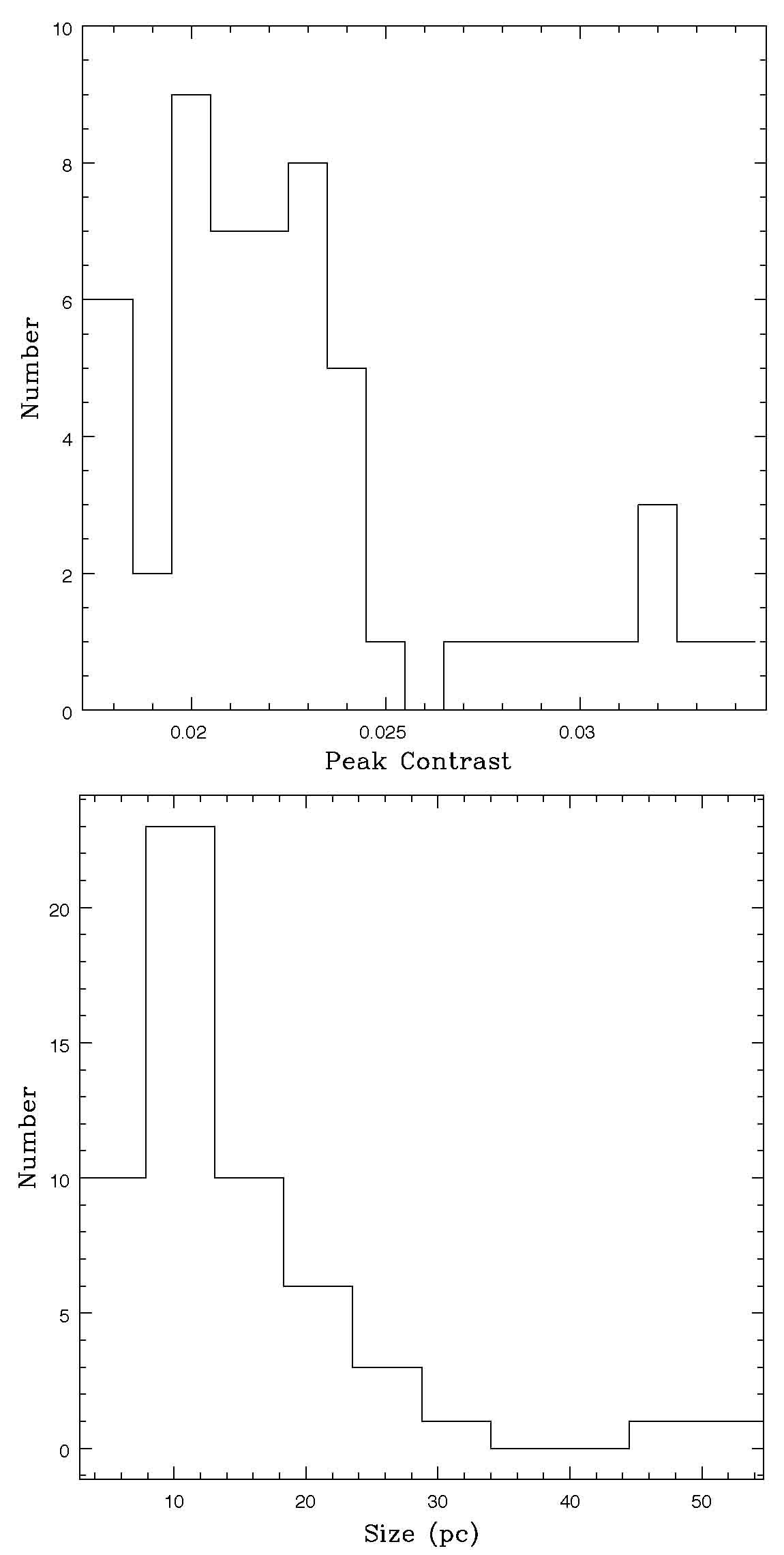}
\vskip 0.3 cm
\begin{minipage}{16cm}
\begin{footnotesize}
\textsc{Fig. 4---} (\textit{Top}) The histogram of peak-contrasts of selected  
55 HCRs. Signal-to-noise ratio $> 5$ was used for the selection. 
(\textit{Bottom}) The histogram of sizes of selected 55 HCRs.
Regions larger than 3 pc were identified as HCRs. 
\end{footnotesize}
\end{minipage}
\end{center}
\end{figure*}

\begin{figure*}
\begin{center}
\includegraphics[scale=0.7]{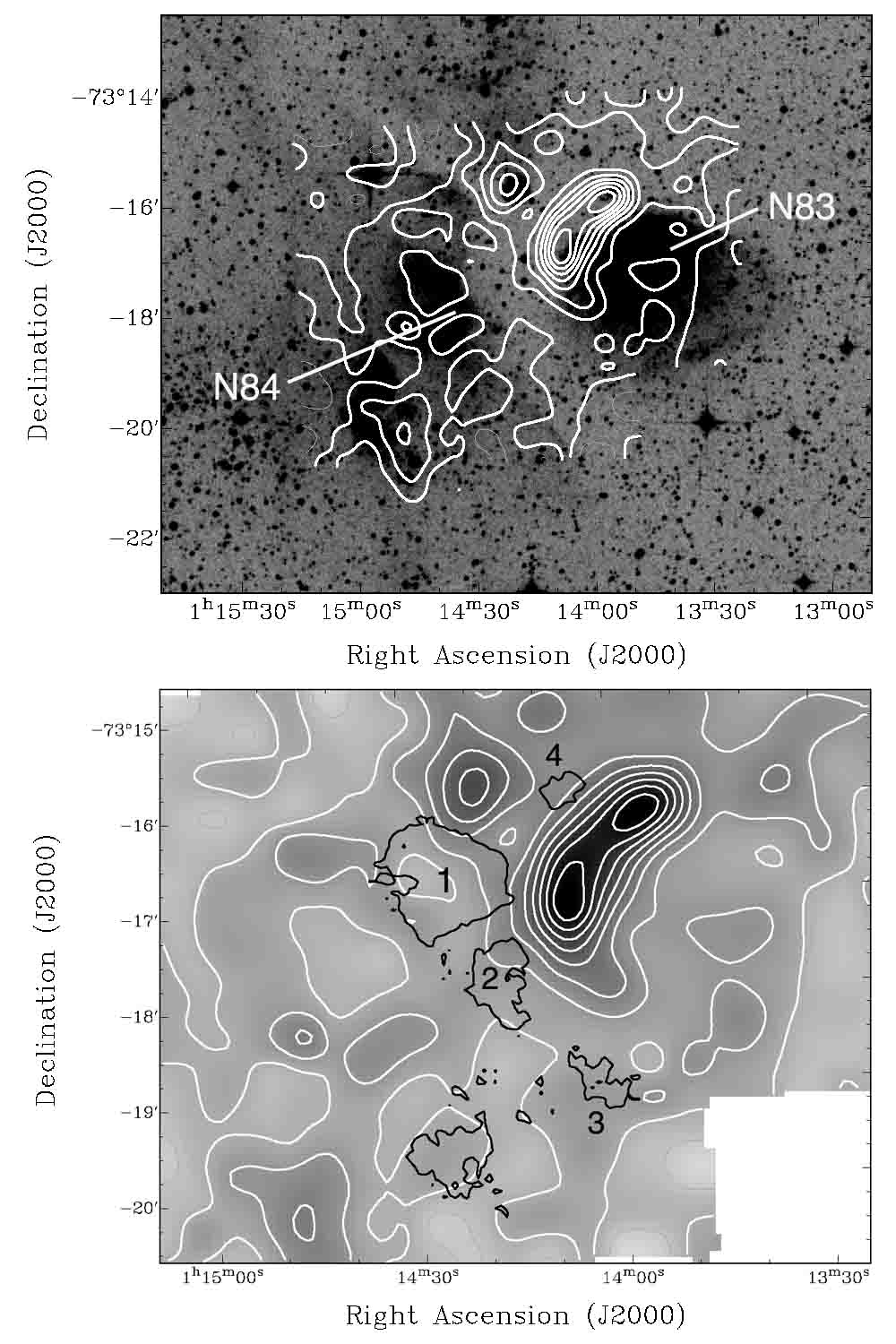}
\vskip 0.3 cm 
\begin{minipage}{16cm}
\begin{footnotesize}
\textsc{Fig. 5---} (\textit{Top}) The CO(2--1) integrated intensity map (Bolatto et al. 2003) 
overlaid on the DSS R band image of the N83/N84 region. 
The CO contour levels are from 10 to 90 \% of the peak 
($\rm 0.0045~K~km~s^{-1}$), with a 10 \% step. 
(\textit{Bottom}) The 4 highest contrast features (from 1 to 4 in order of size) 
are overlaid in black contours on the CO(2--1) integrated intensity map. 
\end{footnotesize}
\end{minipage}
\end{center}
\end{figure*}

\begin{figure*}
\begin{center}
\includegraphics[scale=0.75]{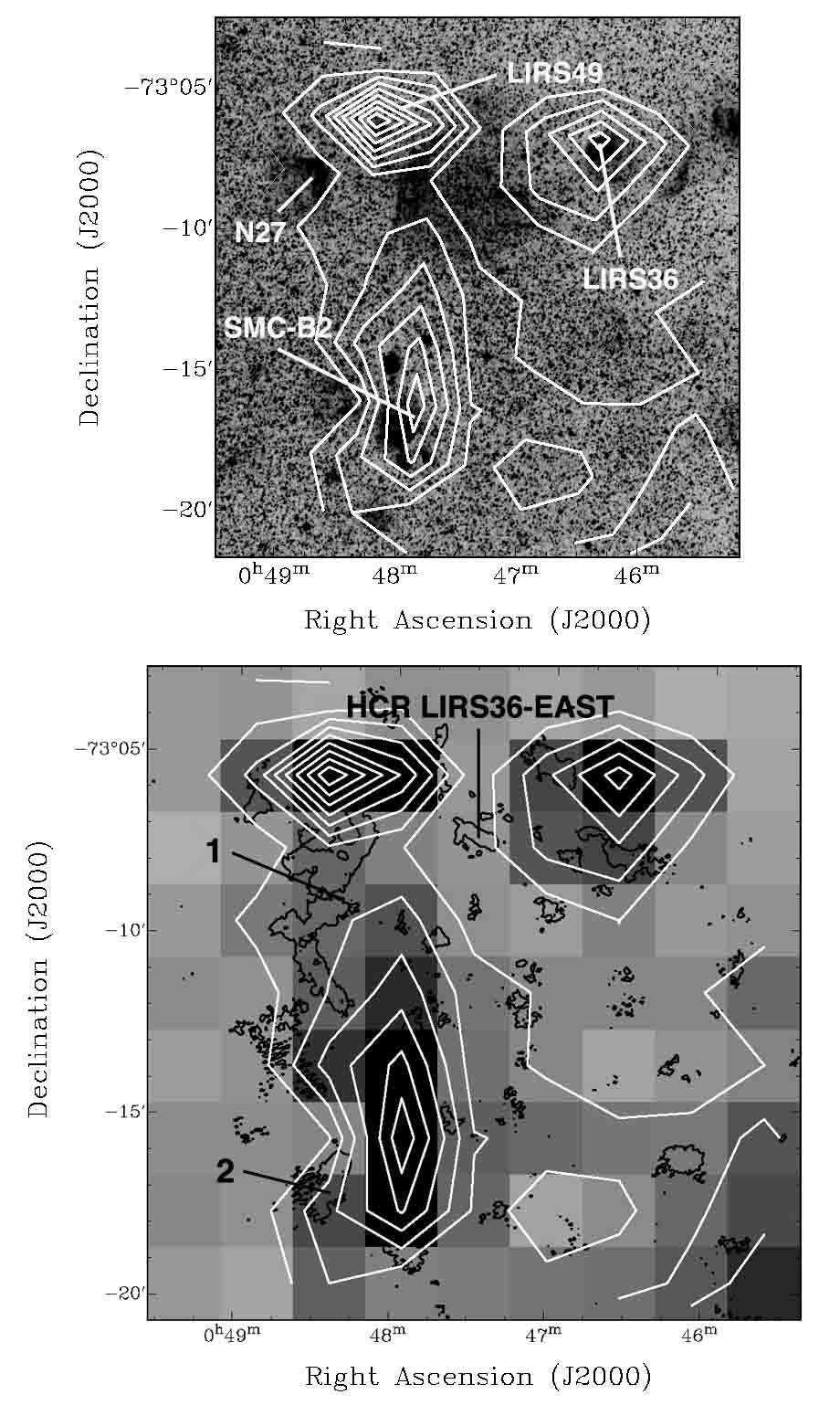}
\vskip 0.3 cm 
\begin{minipage}{16cm}
\begin{footnotesize}
\textsc{Fig. 6---} (\textit{Top}) The CO(1--0) integrated intensity map (Mizuno et al. 2001) 
overlaid on the DSS R band image of the SMC--B2 region. 
The CO contour levels are from 10 to 90 \% of the peak ($\rm 2.49~K~km~s^{-1}$), 
with a 10 \% step.  
Molecular clouds, LIRS36, LIRS49, SMC--B2, and the star-forming region N27 
are labeled. 
(\textit{Bottom}) HCR LIRS36--east and the 2 highest contrast HCRs
(from 1 to 2 in order of size) overlaid in black contours on the CO(1--0) integrated intensity map.
\end{footnotesize}
\end{minipage}
\end{center}
\end{figure*}

\begin{figure*}
\begin{center}
\includegraphics[scale=0.7]{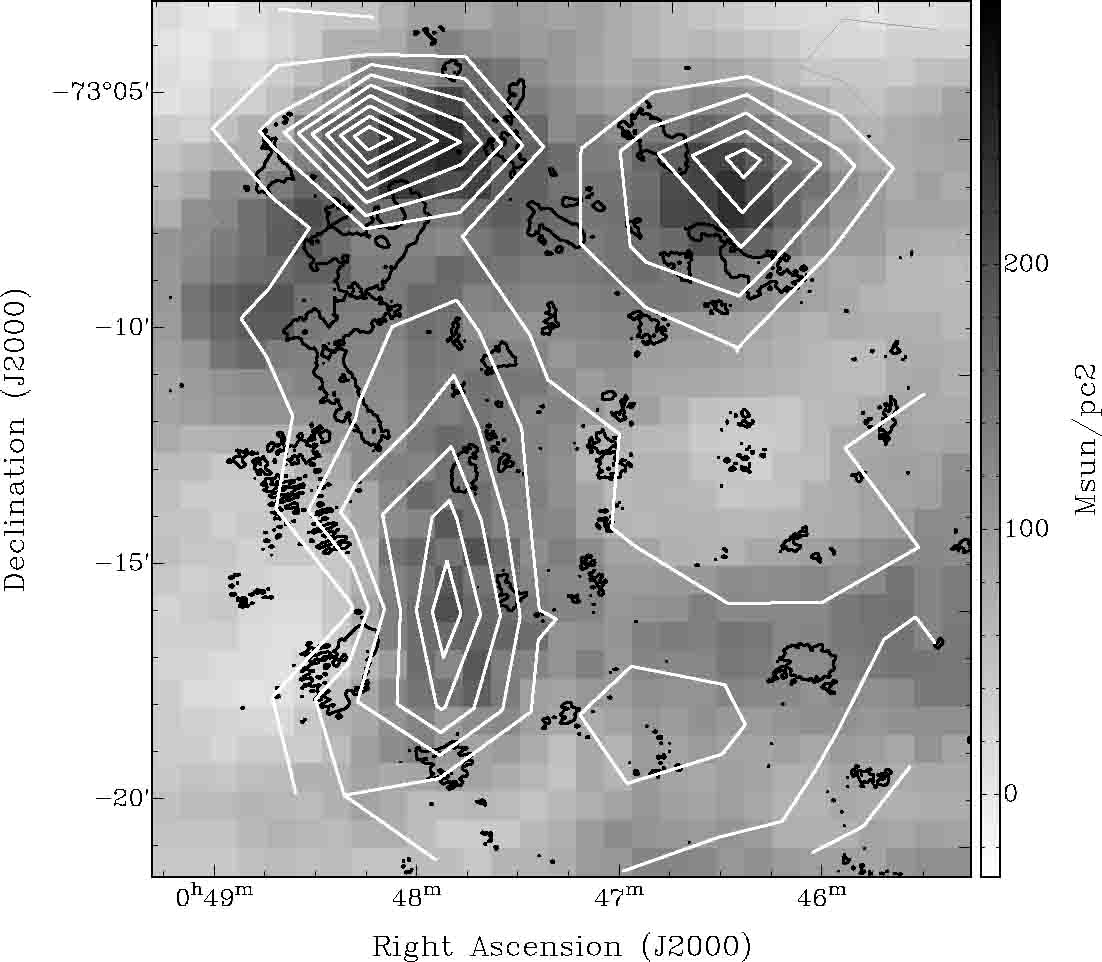}
\vskip 0.3 cm 
\begin{minipage}{16cm}
\begin{footnotesize}
\textsc{Fig. 7---} 
The selected HCRs in the SMC--B2 region 
overlaid in black contours on the H$_{2}$ surface density map of Leroy et al. (2007). 
CO(1--0) emission from the NANTEN survey (Mizuno et al. 2001) 
is also overlaid in white contours. 
The CO contours are from 10 to 90 \% of the peak 
($\rm 2.49~K~km~s^{-1}$), with 10 \% step.
\end{footnotesize}
\end{minipage}
\end{center}
\end{figure*}

\begin{figure*}
\begin{center}
\includegraphics[scale=0.65]{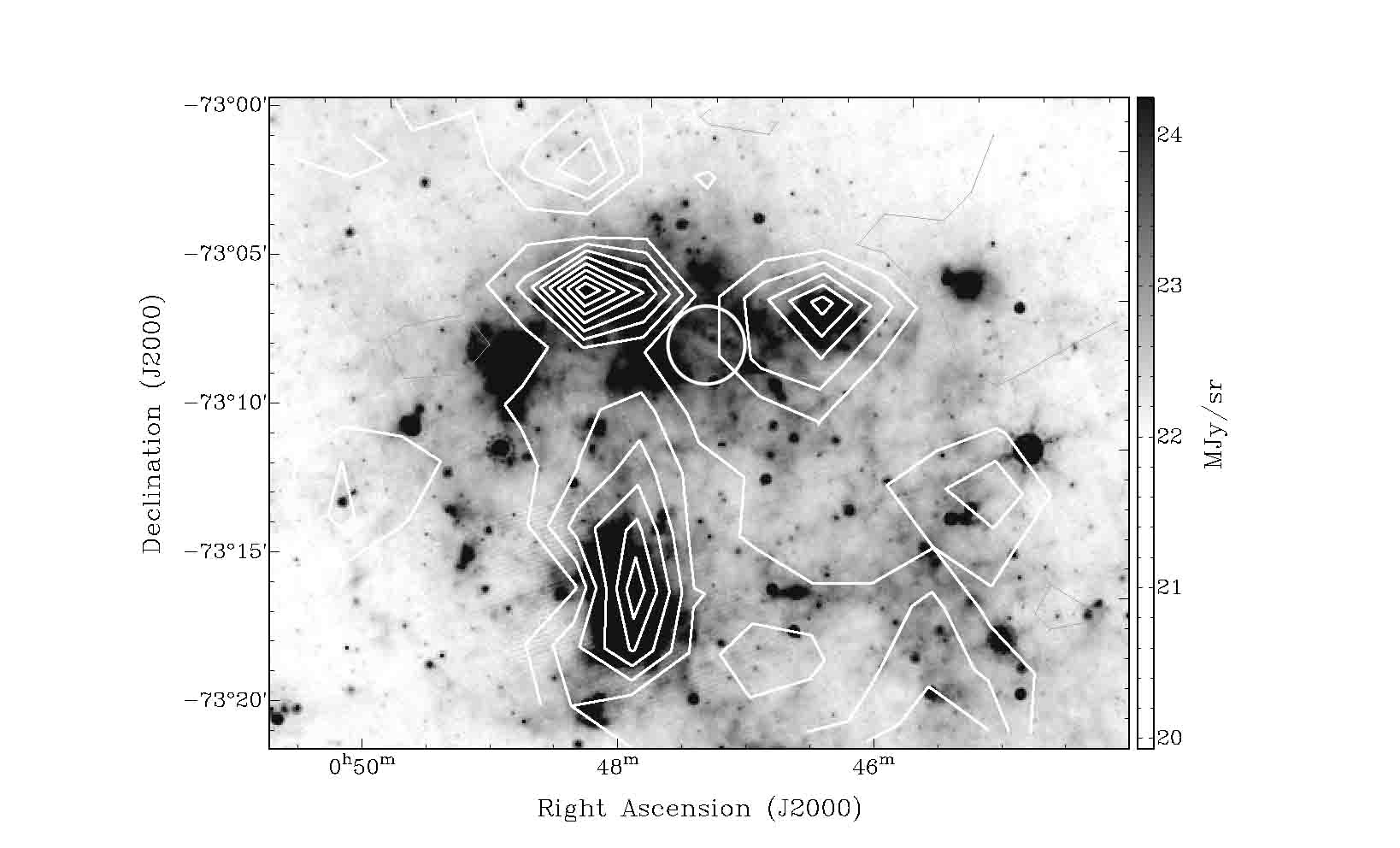}
\vskip 0.3 cm 
\begin{minipage}{16cm}
\begin{footnotesize}
\textsc{Fig. 8---} The \textit{Spitzer} 24 $\mu$m image of the SMC SW bar region.
The intensity scale of the image was stretched to emphasize the location of 
HCR LIRS36-east between molecular clouds LIRS36 and LIRS49.  
The white contours are from the NANTEN CO(1--0) Survey (Mizuno et al. 2001). 
The contour levels are from 10 to 90 \% of the peak ($\rm 2.49~K~km~s^{-1}$), 
with 10 \% step. The primary beam of the ATCA ($2.4'$ at 23 GHz) is centered on
HCR LIRS36--east, 
($\alpha$, $\delta$)$_{\rm J2000}= ($$\rm 00^{h}$:$\rm 47^{m}$:$\rm 30^{s}$, 
$\rm -73^{\circ}$:$\rm 07'$:$\rm 30''$), 
and is shown as a white circle. 
\end{footnotesize}
\end{minipage}
\end{center}
\end{figure*}

\begin{figure*}
\begin{center}
\includegraphics[scale=0.6]{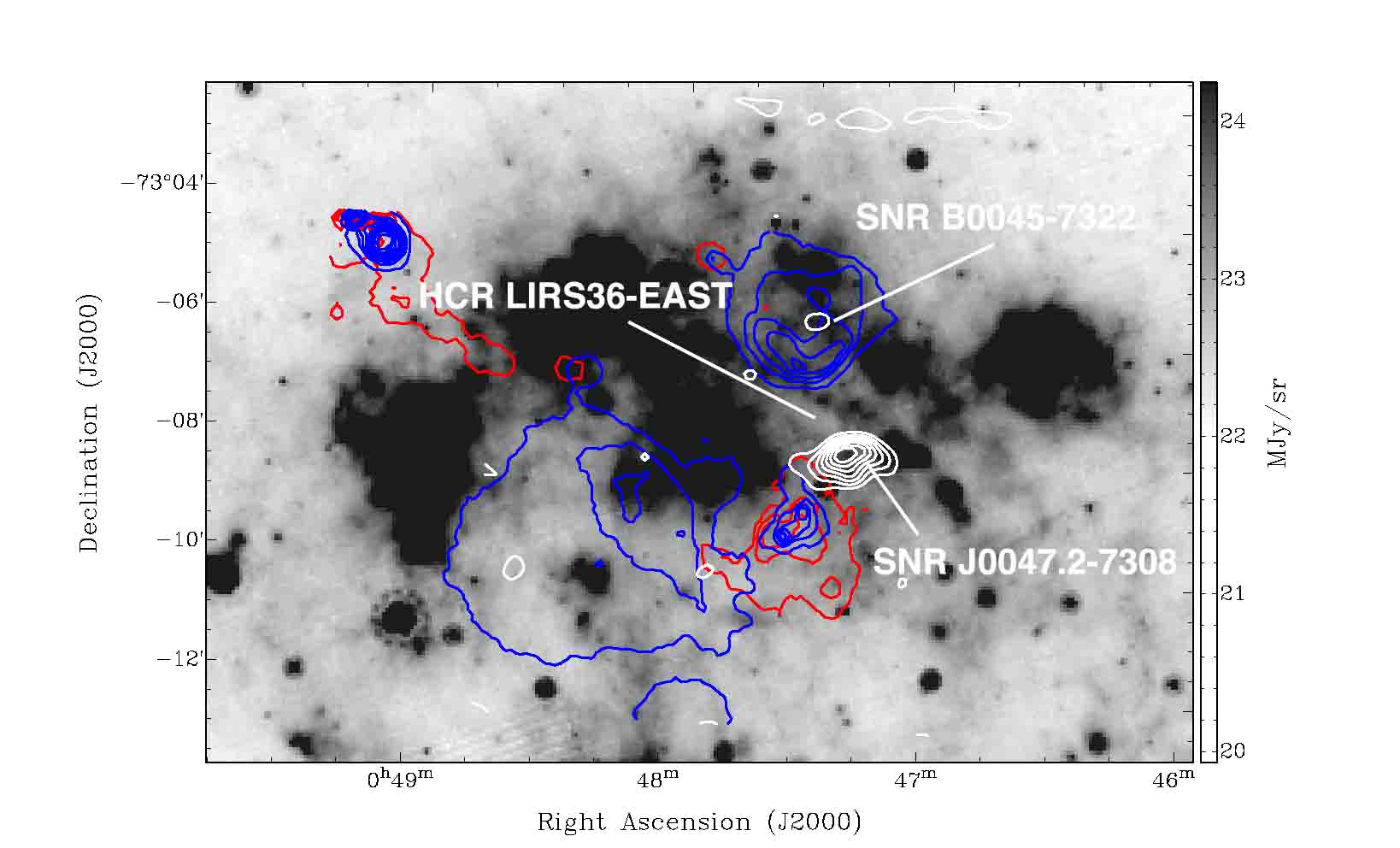}
\vskip 0.3 cm 
\begin{minipage}{16cm}
\begin{footnotesize}
\textsc{Fig. 9---} Emission of $\rm H_{2}$ at 28.2 $\mu$m (red), 
S \textsc{iii} at 33.5 $\mu$m (blue), and 1.2 cm radio continuum (white) overlaid on 
the \textit{Spitzer} 24 $\mu$m image of the SMC SW bar region. 
$\rm H_{2}$ and S \textsc{iii} data were obtained with the IRS (Bolatto et al., in preparation). 
The contour levels range from 
20 to 90 \% of the peak ($\rm 0.16~erg~sec^{-1}~cm^{-2}$)
for $\rm H_{2}$, from 10 to 90 \% of the peak ($\rm 2.52~erg~sec^{-1}~cm^{-2}$)
for S \textsc{iii}, and from 30 to 90 \% of the peak ($\rm 0.0023~mJy~Beam^{-1}$) 
for 1.2 cm emission. All contours have a step of 10 \%. 
\end{footnotesize}
\end{minipage}
\end{center}
\end{figure*}


\begin{figure*}
\begin{center}
\includegraphics[scale=0.6]{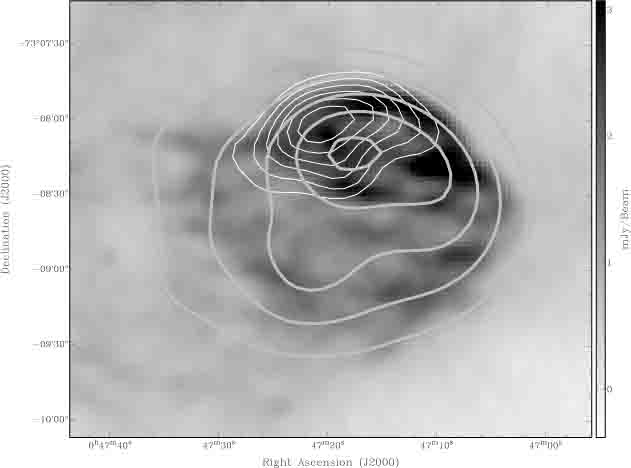}
\vskip 0.3 cm 
\begin{minipage}{16cm}
\begin{footnotesize}
\textsc{Fig. 10---} The 1.34 GHz image of SNR J0047.2--7308. 
4.8 GHz (thick-grey) and 1.2 cm (white) emission are overlaid on the image. 
Images at 1.34 GHz and 4.8 GHz were kindly provided by John Dickel.
The contours range from 5 and 90 \% of the peak ($\rm 0.2~mJy~Beam^{-1}$) with a step of 2 \%
for the 4.8 GHz emission, and  from 40 to 90 \% of the peak 
($\rm 0.0023~mJy~Beam^{-1}$) with a step of 10 \% for
the 1.2 cm emission.
The contours of the 23 GHz (1.2 cm) source follow well the morphology of the
SNR at 1.34 GHz and 4.8 GHz, however its center is off-set slightly from the 
center of the northern rim of the SNR. 
\end{footnotesize}
\end{minipage}
\end{center}
\end{figure*}

\begin{figure*}
\begin{center}
\includegraphics[scale=0.6]{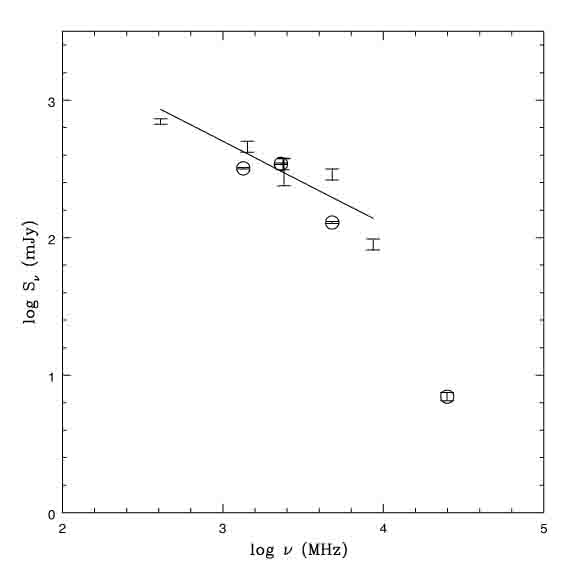}
\vskip 0.3 cm 
\begin{minipage}{16cm}
\begin{footnotesize}
\textsc{Fig. 11---} The radio spectrum of SNR J0047.2--7308: 
 $S_{\nu}$ (flux density) in mJy versus $\nu$ (frequency) in MHz. 
The flux densities at 408 MHz, 1.42 GHz, 2.37 GHz, 2.4 GHz, 4.8 GHz, and 8.64 GHz
of the SNR were collected from the literature and are shown with 1$\sigma$ error bars. 
Our measurements at 1.34 GHz, 4.8 GHz, and 23 GHz are shown as circles, and
encompassed only the portion of the SNR seen at 23 GHz. 
The spectral index at the lower frequencies (solid line) $\alpha=-0.6 \pm 0.2$ 
was estimated using the data from literature.  
\end{footnotesize}
\end{minipage}
\end{center}
\end{figure*}

\end{document}